\documentclass[twocolumn]{aastex631}

\shorttitle{Period-Luminosity Relations of Type II Cepheids}
\shortauthors{Wielg\'orski et al.}

\graphicspath{{./}{}}

\begin{document}

\title{An absolute calibration of the near-infrared Period-Luminosity Relations of Type II Cepheids in the Milky Way and in the Large Magellanic Cloud.}

\author{Piotr Wielg\'orski}
\affiliation{Nicolaus Copernicus Astronomical Center, Polish Academy of Sciences, Bartycka 18, 00-716 Warszawa, Poland}
\email{pwielgor@camk.edu.pl}

\author{Grzegorz Pietrzy\'nski}
\affiliation{Nicolaus Copernicus Astronomical Center, Polish Academy of Sciences, Bartycka 18, 00-716 Warszawa, Poland}
\affiliation{Universidad de Concepci\'on, Departamento de Astronomia, Casilla 160-C, Concepci\'on, Chile}

\author{Bogumi\l{} Pilecki}
\affiliation{Nicolaus Copernicus Astronomical Center, Polish Academy of Sciences, Bartycka 18, 00-716 Warszawa, Poland}

\author{Wolfgang Gieren}
\affiliation{Universidad de Concepci\'on, Departamento de Astronomia, Casilla 160-C, Concepci\'on, Chile}

\author{Bart\l{}omiej Zgirski}
\affiliation{Nicolaus Copernicus Astronomical Center, Polish Academy of Sciences, Bartycka 18, 00-716 Warszawa, Poland}

\author{Marek G\'orski}
\affiliation{Nicolaus Copernicus Astronomical Center, Polish Academy of Sciences, Bartycka 18, 00-716 Warszawa, Poland}

\author{Gergely Hajdu}
\affiliation{Nicolaus Copernicus Astronomical Center, Polish Academy of Sciences, Bartycka 18, 00-716 Warszawa, Poland}

\author{Weronika Narloch}
\affiliation{Universidad de Concepci\'on, Departamento de Astronomia, Casilla 160-C, Concepci\'on, Chile}

\author{Paulina Karczmarek}
\affiliation{Universidad de Concepci\'on, Departamento de Astronomia, Casilla 160-C, Concepci\'on, Chile}

\author{Rados\l{}aw Smolec}
\affiliation{Nicolaus Copernicus Astronomical Center, Polish Academy of Sciences, Bartycka 18, 00-716 Warszawa, Poland}

\author{Pierre Kervella}
\affiliation{LESIA, Observatoire de Paris, Universit'e PSL, CNRS, Sorbonne Universit'e, Universit'e de Paris, 5 place Jules Janssen, 92195 Meudon,France}

\author{Jesper Storm}
\affiliation{Leibniz-Institut f\"ur Astrophysik Potsdam (AIP), An der Sternwarte 16, 14482 Potsdam, Germany}

\author{Alexandre Gallenne}
\affiliation{Universidad de Concepci\'on, Departamento de Astronomia, Casilla 160-C, Concepci\'on, Chile}
\affiliation{Unidad Mixta Internacional Franco-Chilena de Astronomía (CNRS UMI 3386), Departamento de Astronomía, Universidad deChile, Camino El Observatorio 1515, Las Condes, Santiago, Chile}

\author{Louise Breuval}
\affiliation{LESIA, Observatoire de Paris, Universit'e PSL, CNRS, Sorbonne Universit'e, Universit'e de Paris, 5 place Jules Janssen, 92195 Meudon,France}
\affiliation{Department of Physics and Astronomy, Johns Hopkins University, Baltimore, MD 21218, USA}

\author{Megan Lewis}
\affiliation{Nicolaus Copernicus Astronomical Center, Polish Academy of Sciences, Bartycka 18, 00-716 Warszawa, Poland}

\author{Miko\l{}aj Ka\l{}uszy\'nski}
\affiliation{Nicolaus Copernicus Astronomical Center, Polish Academy of Sciences, Bartycka 18, 00-716 Warszawa, Poland}

\author{Dariusz Graczyk}
\affiliation{Nicolaus Copernicus Astronomical Center, Polish Academy of Sciences, Rabia\'nska 8, 87-100, Toru\'n, Poland}

\author{Wojciech Pych}
\affiliation{Nicolaus Copernicus Astronomical Center, Polish Academy of Sciences, Bartycka 18, 00-716 Warszawa, Poland}

\author{Ksenia Suchomska}
\affiliation{Nicolaus Copernicus Astronomical Center, Polish Academy of Sciences, Bartycka 18, 00-716 Warszawa, Poland}

\author{M\'onica Taormina}
\affiliation{Nicolaus Copernicus Astronomical Center, Polish Academy of Sciences, Bartycka 18, 00-716 Warszawa, Poland}

\author{Gonzalo Rojas Garcia}
\affiliation{Nicolaus Copernicus Astronomical Center, Polish Academy of Sciences, Bartycka 18, 00-716 Warszawa, Poland}

\author{Aleksandra Kotek}
\affiliation{Nicolaus Copernicus Astronomical Center, Polish Academy of Sciences, Bartycka 18, 00-716 Warszawa, Poland}

\author{Rolf Chini}
\affiliation{Astronomisches Institut, Ruhr-Universit\"at Bochum, Universit\"atsstrasse 150, D-44801 Bochum, Germany}
\affiliation{Instituto de Astronom\'{i}a, Universidad Cat\'{o}lica del Norte, Avenida Angamos 0610, Antofagasta, Chile}
\affiliation{Nicolaus Copernicus Astronomical Center, Polish Academy of Sciences, Bartycka 18, 00-716 Warszawa, Poland}

\author{Francisco Pozo Nu\~nez}
\affiliation{Astronomisches Institut, Ruhr-Universit\"at Bochum, Universit\"atsstrasse 150, D-44801 Bochum, Germany}
\affiliation{ Astroinformatics, Heidelberg Institute for Theoretical Studies, Schloss-Wolfsbrunnenweg 35, 69118 Heidelberg, Germany}

\author{Sadegh Noroozi}
\affiliation{Astronomisches Institut, Ruhr-Universit\"at Bochum, Universit\"atsstrasse 150, D-44801 Bochum, Germany}

\author{Catalina Sobrino Figaredo}
\affiliation{Astronomisches Institut, Ruhr-Universit\"at Bochum, Universit\"atsstrasse 150, D-44801 Bochum, Germany}

\author{Martin Haas}
\affiliation{Astronomisches Institut, Ruhr-Universit\"at Bochum, Universit\"atsstrasse 150, D-44801 Bochum, Germany}

\author{Klaus Hodapp}
\affiliation{University of Hawaii, Institute for Astronomy, 640 N. Aohoku Place, Hilo, HI 96720, USA}

\author{Przemys\l{}aw Miko\l{}ajczyk}
\affiliation{Astronomical Institute, University of Wroc\l{}aw, M. Kopernika 11, 51–622 Wroc\l{}aw, Poland}

\author{Krzysztof Kotysz}
\affiliation{Astronomical Institute, University of Wroc\l{}aw, M. Kopernika 11, 51–622 Wroc\l{}aw, Poland}

\author{Dawid Mo\'zdzierski}
\affiliation{Astronomical Institute, University of Wroc\l{}aw, M. Kopernika 11, 51–622 Wroc\l{}aw, Poland}

\author{Piotr Ko\l{}aczek-Szyma\'nski}
\affiliation{Astronomical Institute, University of Wroc\l{}aw, M. Kopernika 11, 51–622 Wroc\l{}aw, Poland}

\begin{abstract}

We present time-series photometry of 21 nearby Type II Cepheids in the near-infrared $J$, $H$ and $K_s$ passbands. We use this photometry, together with the Third $Gaia$ Early Data Release parallaxes, to determine for the first time period-luminosity relations (PLRs) for Type II Cepheids from field representatives of these old pulsating stars in the near-infrared regime. We found PLRs to be very narrow for BL Herculis stars, which makes them candidates for precision distance indicators. We then use archival photometry and the most accurate distance obtained from eclipsing binaries to recalibrate PLRs for Type II Cepheids in the Large Magellanic Cloud (LMC). Slopes of our PLRs in the Milky Way and in the LMC differ by slightly more than $2\sigma$ and are in a good agreement with previous studies of the LMC, Galactic Bulge and Galactic Globular Clusters Type II Cepheids samples. We use PLRs of Milky Way Type II Cepheids to measure the distance to the LMC and we obtain a distance modulus of 18.540$\pm$0.026(stat.)$\pm$0.034(syst.)mag in the $W_{JK}$ Wesenheit index. We also investigate the metallicity effect within our Milky Way sample and we find rather significant value of about -0.2mag/dex in each band meaning that more metal-rich Type II Cepheids are intrinsically brighter than their more metal-poor counterparts, in agreement with the value obtained from Type II Cepheids in Galactic Globular Clusters. The main source of systematic error on our Milky Way PLRs calibration and the LMC distance is the current uncertainty of the $Gaia$ parallax zero point.
\end{abstract}


\section{Introduction} \label{sec:intro}
The distinction between Classical Cepheids (CCeps) and Type II Cepheids (T2Ceps) \citep{1944ApJ...100..137B, 1958AJ.....63..207B} led to the revision of the cosmic distance scale and significantly increased the measured distances to galaxies and, by extension, the timescale of the Universe. Mixing these two classes of pulsating stars for the distance determination using the Leavitt Law \citep[Period-Luminosity Relation, PLR][]{1908AnHar..60...87L} as a standard candle results in a significant inaccuracy as T2Ceps are 1.5-2 magnitudes fainter than CCeps of similar periods. CCeps remain the most important distance indicators, acting as an anchor for the local cosmic distance scale and determination of the Hubble Constant \citep{2021ApJ...908L...6R}. Less luminous T2Ceps are not as famous, however, they are crucial objects for tracing old stellar populations in our and nearby galaxies and globular clusters \citep{2018A&A...619A..51B}. Their lower brightness and abundance \citep[$\sim$10000 CCeps and only $\sim$300 T2Ceps are observed in the Magellanic Clouds][]{2017AcA....67..103S,2018AcA....68...89S} make them more difficult to observe in other galaxies and to apply them as extragalactic distance indicators. However, there are objects like e.g. globular clusters or dwarf spheroidal galaxies in which CCeps are not observed at all and T2Ceps can be applied as distance tracers in these instances, more effectively than RR Lyrae stars as T2Ceps are 1-4 magnitude brighter.

T2Ceps are divided based on their period distributions \citep{1985MmSAI..56..169G} into three subgroups: BL Herculis (BL Her) type stars with periods between $\sim$1 and $\sim$5 days, W Virginis (W Vir) type stars with periods between $\sim$5 and $\sim$20 days and RV Tauri (RV Tau) stars with periods longer than $\sim$20 days. This distinction is not strict and depends on the environement. Stars belonging to each group present different light curve morphology and are most probably at different evolutionary stages. \citet{2008AcA....58..293S} separated another subgroup of T2Ceps in the Magellanic Clouds which are usually called peculiar W Virginis (pW Vir, pWV) stars as they cover a similar range of periods as "classical" W Vir stars but are usually brighter, bluer and show different light curve morphology from W Vir stars. Many of these peculiar stars were found to be in binary systems \citep{2017AcA....67..297S}.  

Evolutionary channels leading to formation of T2Ceps are not fully explained but these sources are low metallicity ([Fe/H] between $\sim $-2.5 and $\sim $0 dex) and most probably low mass (0.5-1$M_{\odot}$) stars crossing the Instability Strip during transition from the blue horizontal branch to asymptotic giant branch in the case of BL Her stars, as a result of helium-shell-flashes for W Vir stars and post asymptotic giant branch in the case of RV Tau stars \citep{1997A&A...317..171B, 2002PASP..114..689W}. Detailed studies of two pW Vir stars in the Large Magellanic Cloud (LMC) resulted in the first dynamical mass determination for T2Ceps \citep[0.64$\pm$0.02 and 1.51$\pm$0.09 $M_{\odot}$]{2017ApJ...842..110P,2018ApJ...868...30P}. \citet{2018ApJ...868...30P} conclude that pW Vir stars are products of binary evolution \citep[similar to Binary Evolution Pulsators][]{2012Natur.484...75P} and they are much younger than other T2Ceps. Recent studies of the spatial distribution of T2Ceps in the Magellanic Clouds \citep{2018AcA....68..213I} as well as infrared excess in the Spectral Energy Distribution and the Light-time Travel Effect search in T2Ceps light curves \citep{2017A&A...603A..70G} suggest that the W Vir subgroup might be a mixture of old and intermediate stellar populations and it is possible that these stars are result of the binary evolution. Amplitude and period variations are common for T2Ceps \citep{2010AJ....139.2300R, 2016JAVSO..44..179N}, particularly among W Vir and RV Tau type stars, which can be partially explained by their evolution and movement across the Intability Strip \citep{2016JAVSO..44..179N} but binarity, period modulations, period doubling and chaotic dynamics also contribute \citep{1990ApJ...355..590M,2016MNRAS.456.3475S,2017MNRAS.465..173P,2018MNRAS.481.3724S}. Almost all known T2Ceps pulsate in the fundamental mode \citep{1997A&A...317..171B}. Two first-overtone stars were found among the LMC T2Ceps by  \citet{2019ApJ...873...43S}.

T2Ceps could be useful for determining distances of galaxies up to several megaparsecs using existing instruments \citep{2009AcA....59..403M}. PLRs of T2Ceps have been investigated empirically in the optical and near-infrared domain by many authors in the neighbouring Large and Small Magellanic Clouds \citep{1998AJ....115.1921A,2011MNRAS.413..223M, 2010AcA....60..233C, 2015MNRAS.446.3034R, 2017AJ....153..154B, 2018AcA....68...89S},  Galactic Globular Clusters (GGCs) \citep{2006MNRAS.370.1979M} and Galactic Bulge (GB) \citep{2011AcA....61..285S, 2018A&A...619A..51B} while theoretical work is so far very limited compared to CCeps or RR Lyrae stars \citep{2007A&A...471..893D, 2021MNRAS.501..875D}. \citet{2019A&A...625A..14R} provided PLRs for T2Ceps in the solar neighborhood in the $Gaia$ optical bands using parallaxes from the $Gaia$ Data Release 2. According to these studies period distributions vary between systems and PLRs are linear for BL Her and W Vir stars, while steepening for RV Tau stars. Non-negligible metallicity effect on the absolute brightness of T2Ceps was found by \citet{2006MNRAS.370.1979M} and by \citet{2021MNRAS.501..875D} but with opposite signs. \citet{2006MNRAS.370.1979M} also suggested, that RR Lyrae variables and T2Ceps create a common linear PLR in the near-infrared regime. The zero-point of the T2Cep distance scale was calibrated by \citet{2008MNRAS.386.2115F} based on the Hipparcos parallaxes of two nearby stars- VY Pyx and $\kappa$ Pav and pulsational parallaxes of $\kappa$ Pav, V553 Cen and SW Tau. 

As radially pulsaing stars, T2Ceps also offer the possibility of measuring semi-geometrical distances with the Baade-Wesselink technique \citep{2008MNRAS.386.2115F}, but studies of their projection factor (translating apparent radial velocity into the velocity of pulsating surface of the star) are limited just to a few pioneering works \citep[e.g.][]{1997AJ....113.1833B,2015A&A...576A..64B,2017ApJ...842..110P}. A very good summary of our current knowledge about T2Ceps and efforts made to use them as distance indicators can be found in the recent review papers by \citet{2018SSRv..214..113B} and \citet{2020JApA...41...23B}. 

Currently, the $Gaia$ space mission \citep{2016A&A...595A...1G} provides an unprecedented 6-dimensional map of our stellar neighbourhood. This gives us an opportunity to use nearby stars of the general field with parallaxes known to within 1\% accuracy to set the zero point of the cosmic distance scale with similar accuracy. In this study we present the first calibration of PLRs of T2Ceps in the Milky Way (MW) using near-infrared time-series photometry and the Third $Gaia$ Early Data Release \citep[EDR3][]{2020yCat.1350....0G} parallaxes of nearby representatives of this class of pulsating stars. Presented work opens a series of publications of the Araucaria Project \citep{2002AJ....123..789P, 2005ApJ...628..695G} regarding calibration of different distance indicators using multiband time-series photometry and spectroscopy of nearby stars and $Gaia$ parallaxes. The Araucaria project was started in 2002 and our main motivation is precision calibration of the cosmic distance scale and determination of the local value of the Hubble Constant ($H_0$) with 1\% accuracy. Cosmic distances and, by extension, $H_0$ uncertainty are dominated by systematic errors and high accuracy can be achieved only by using several independent and precise types of distance indicators to determine distance to a given object and compare results. 

The publication is organized as follows: in section \ref{sec:data} we describe our observations, photometry and determination of absolute mean magnitudes of our target stars in the MW as well as preparation of the LMC photometry. Section \ref{sec:results} presents calibration of PLRs, investigation of the effect of metalicity on T2Ceps magnitudes and measurement of the distance to the LMC. We discuss our results in section \ref{sec:disc} and summarize our work and prospects for the future in section \ref{sec:concl}.

\section{Data} \label{sec:data}

\subsection{The Milky Way} \label{subsec:mwdata}

\subsubsection{Sample selection}

\begin{widetext}

\begin{figure*}[t!]
\epsscale{1.1}
\plotone{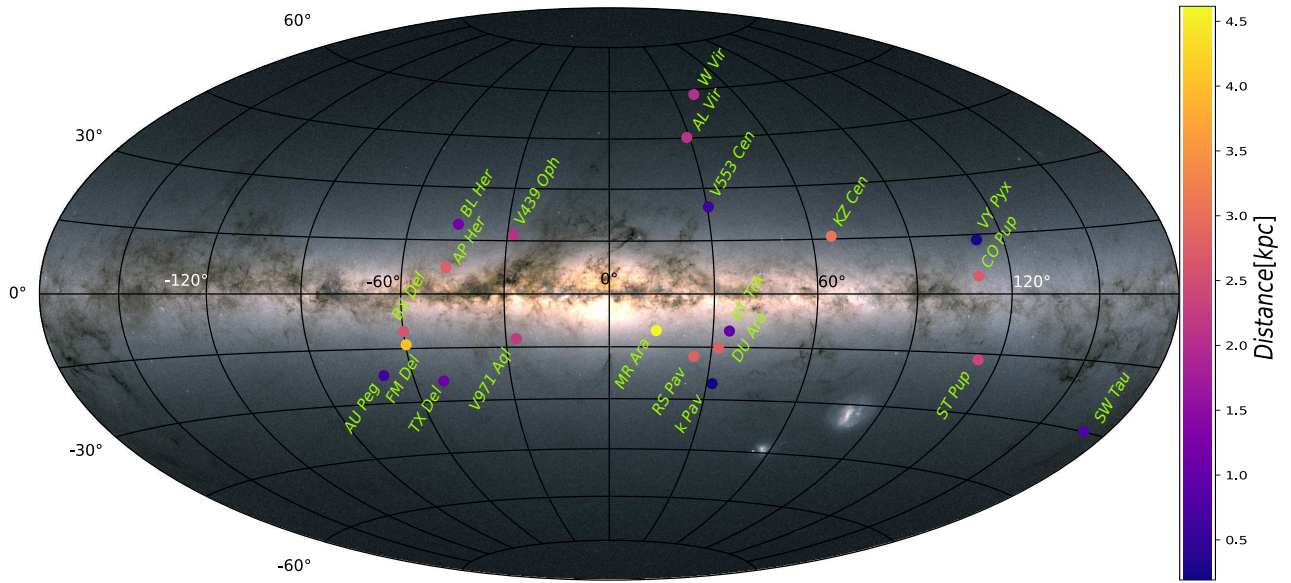}
\caption{The $Gaia$ photometric map of the Milky Way with marked positions of T2Ceps considered in this work. Color denotes the distance of a given star from the Sun.\label{fig:mw_map}}
\end{figure*}

\end{widetext}

From the SIMBAD \footnote{\url{http://simbad.u-strasbg.fr/simbad/}} and AAVSO \footnote{\url{https://www.aavso.org/vsx/index.php?view=search.top}} databases we selected stars classified as T2Ceps with estimated distances from previous studies smaller than 5kpc from the Sun as the expected accuracy of $Gaia$ parallaxes for stars at such distances is better than 5$\%$. At such level of precision, bias on absolute magnitudes inferred from parallaxes should be negligible \citep[see e.g.][]{1973PASP...85..573L, 1997MNRAS.286L...1F, 1999ASPC..167...13A}. Another selection criterion was the $V$-band magnitude. We discarded stars fainter than 13mag in the $V$-band, as for such stars we would not be able to obtain precision (S/N $\sim$100) near-infrared photometry with a 0.8m telescope. We adopted a boundary period of 20 days to distinguish stars belonging to W Vir and RV Tau subclasses. The latter was not considered as these stars do not obey common linear PLR in both the Large and Small Magellanic Clouds. For 19 stars from this starting sample we collected photometric data in $J$, $H$ and $K_s$ bands closely imitating the Two Micron All Sky Survey (2MASS) system \citep{2006AJ....131.1163S}. Fig. \ref{fig:mw_map} presents positions of stars considered in this study plotted on the $Gaia$ photometric map of the MW.

\subsubsection{NIR observations and data reductions}

Observations were handled between February 2017 and March 2020 using the 0.81m InfraRed Imaging Survey (IRIS) telescope equipped with 1kx1k HAWAII detector (resulting in 7 arcmin x 7 arcmin field of view), located in Observatorio Cerro Armazones (OCA) in Chile \citep{2010SPIE.7735E..1AH,2016SPIE.9911E..2MR}. The saturation limit of IRIS is $\sim$7.8mag in $H$ and $\sim$7.5mag in $J$ and $K_s$ passbands. For one star (VY Pyx) we used a neutral density (ND) filter blocking 97\% of the light. This filter allowed us to observe stars up to $\sim$4mag. The ND filter was not cooled thus it produced thermal radiation reducing the precision of our photometry. Some of our targets were at a saturation limit so we use exposures that were slightly defocused or taken during worse seeing conditions to ensure that they are not saturated. We aimed to cover the full light curve of every star, and in many cases we collected 15-30 data points in each band. In some cases we were only able to collect a few points so far, but our data are uniformly distributed over the pulsation cycle, which allows us to determine mean magnitudes with satisfactory precision without using templates. 

\begin{widetext}

\begin{figure}[t!]
\epsscale{1.}
\plotone{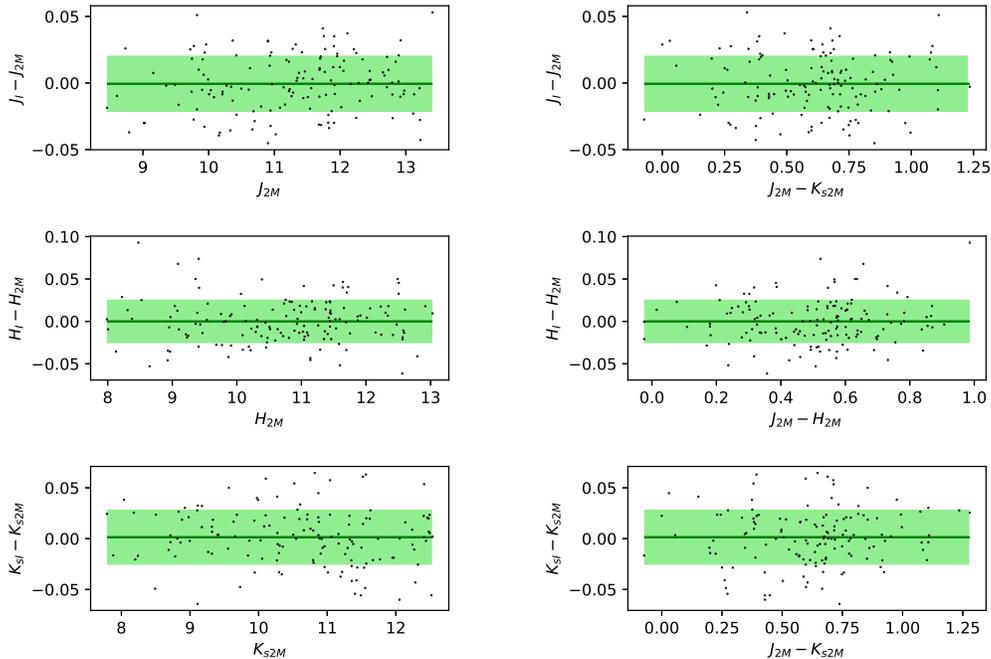}
\caption{Test of the accuracy of our transformations of IRIS photometry from instrumental to the 2MASS system. Magnitudes of constant stars observed by IRIS and transformed to the 2MASS system are compared with corresponding values from the 2MASS catalogue. Typical error in the y axis is 0.03mag (errorbars are not plotted for clarity). Green lines mark the mean difference between our measurement and 2MASS catalogue and it is consistent with zero in all bands with the error on the mean 0.002mag. Light-green area denotes r.m.s., which amounts to 0.022mag in $J$ and 0.025mag in $H$ and $K_s$ passbands.\label{fig:iris_accuracy}}
\end{figure}

\end{widetext}

IRIS always takes short exposure (with the shortest possible exposure time of 2.2s) and long exposure with specified exposure time (14.5s) one by one. In order to estimate and subtract the sky background, each field was observed in 10 dithered positions. Raw exposures were sky-subtracted and flatfielded using standard IRAF \citep{1986SPIE..627..733T} routines. An astrometric solution was performed with Sextractor \citep{1996A&AS..117..393B} and SCAMP \citep{2006ASPC..351..112B} and then 10 dithered frames were resampled and combined with SWARP \citep{2010ascl.soft10068B} into a single final image \citep[details of the pipeline used for calibrations can be found in][]{2012PhDW}. Aperture photometry was performed with the DAOPHOT \citep{1987PASP...99..191S} package and instrumental photometry was tied to the 2MASS system using constant stars present in a given field as standards (usually more than 3 stars of brightness similar to our target with quality flag AAA in the 2MASS catalogue). If there were no comparison stars of similar magnitude in a given field (cases of VY Pyx, SW Tau and AL Vir) we used long exposures (14.5s) to measure the brightness of comparison stars while the target was measured in the short exposure (2.2s). We found a non-negligible color term in $J$ band and it amounts to -0.07$(J-K_s)$. Internal precision of our photometry is at a level of 0.02mag. In order to check correctness of our transformation to the 2MASS system we compared magnitudes of constant stars present in observed fields (transformed to the 2MASS system using an approach identical to our scientific objects) with corresponding magnitudes from the 2MASS catalogue. The observed fields offered very limited numbers of bright constant stars, thus for this test we mostly used sources that were adopted as comparison stars in the transformation of the photometry of science targets. Each considered star was excluded from the set of comparison stars while transforming its own photometry. This test is presented in Fig. \ref{fig:iris_accuracy}. The mean difference between our and 2MASS photometry is zero with the error on the mean of 0.002mag in each band. We adopt this value as our zero-point uncertainty, which contributes to the systematic error of our calibration of PLRs. 

For 3 bright T2Ceps - $\kappa$ Pav, V553 Cen and SW Tau - archival infrared photometry collected at the South African Astronomical Observatory (SAAO) with the MkII photometer installed on the 0.75 metre telescope is available \citep{2008MNRAS.386.2115F}. We transform these photometric data to the 2MASS system using formulae from \citet{2007MNRAS.380.1433K}.

Small fraction of time-series photometry used in this work is presented in Table \ref{tab:t2cep_phot}. Its full version is provided as supplementary material online.

\subsubsection{Optical photometry and periods}

We searched for well-sampled optical photometry for determining periods of T2Ceps. For most of our stars ASAS-SN \citep{2017PASP..129j4502K} and ASAS \citep{2002AcA....52..397P} photometry in V band is available although in some cases light curves were very noisy. Some T2Ceps had been observed by \citet{2008yCat.2285....0B} and we also collected our own optical photometric data in V band with the 0.4m VYSOS 16 telescope located in OCA. VYSOS 16 data were processed using the same procedure as in the case of IRIS, and aperture photometry was performed with DAOPHOT. Then we roughly standardized light curves using a single comparison star and its ASAS-SN magnitude. If quality of ASAS-SN or ASAS light curve was satisfactory we preffered these sources as they contain hundreds of observations while in the case of observations from Berdnikov and VYSOS16 only a few dozen data points were collected. We adopted photometry for 8 stars (BL Her, RT TrA, V553 Cen, AU Peg, AL Vir, CO Pup, W Vir and MR Ara) from ASAS, 9 stars from ASAS-SN (BX Del, KZ Cen, V971 Aql, DU Ara, V439 Oph, FM Del, AP Her, ST Pup and RS Pav), 1 star from Berdnikov ($\kappa$ Pav) and 3 stars from our own VYSOS 16 data (VY Pyx, SW Tau and TX Del).

Primarliy, we adopted the periods available in the AAVSO database and with these initial values we tried to find the value which yielded the smoothest light curve using \texttt{fnpeaks} code \footnote{\url{http://helas.astro.uni.wroc.pl/deliverables.php?active=fnpeaks}}. Fig. \ref{fig:blher_vlc} and \ref{fig:wvir_vlc} present our collection of $V$-band light curves for our sample phased with the final periods and zero-phase set to the intensity-averaged mean value on the rising branch. Periods are given in column 2 of Table \ref{tab:t2cep_data}.

\subsubsection{Mean magnitudes and extinction}

In order to determine mean magnitudes of our infrared light curves we transformed magnitudes into fluxes and modeled each light curve using the following Monte Carlo procedure. For each datapoint of a given lightcurve we generated a random synthetic datapoint from the Gaussian distribution centered on the original datapoint value with the standard deviation equal to the corresponding photometric uncertainty. Then an Akima spline, implemented in the \texttt{Python} \texttt{SciPy} module \citep{2020SciPy-NMeth}, was fitted to the synthetic light curve. Mean flux was calculated and transformed back into magnitude. We repeated this procedure 10,000 times. Fitting a Gaussian to the resulting histogram of mean magnitudes gives us the final mean magnitude and its uncertainty. Our results are given in columns 10-15 of Table \ref{tab:t2cep_data} and phased light curves are presented in Fig. \ref{fig:blher_jlc}-\ref{fig:wvir_klc}. SW Tau is the only star which has observations both from IRIS and MkII photometer and obtained light curves are in a very good agreement. 

Apparent mean magnitudes have to be corrected for interstellar extinction. We used \citet{2011ApJ...737..103S} reddening maps to obtain the color excess $E(B-V)$ in the direction of each star. For the nearest stars in our sample such reddening can be overestimated so we implement the following procedure from \citet{2001ApJ...556..181D}. Assuming an axisymmetric distribution of the dust density in the Galaxy:
\begin{equation}
\rho(r,z) = \rho_0 \exp (-r/r_D - |z|/z_D)  
\end{equation}
where $r$ is the radial coordinate with respect to the galactic center and $z$ is the distance from the galactic plane, we estimated the contribution of dust located between us and the target star to the total $E(B-V)$ value in a given direction. Detailed description of the model used in this process can be found in \citet{2015MNRAS.451..651S}. Final values of $E(B-V)$ for our sample stars are given in the column 17 of Table \ref{tab:t2cep_data} and in most cases they do not differ significantly from original values from \citet{2011ApJ...737..103S} map. Adopting the reddening law from \citet{1989ApJ...345..245C} and \citet{1994ApJ...422..158O} and $R_V=3.1$ we calculated total-to-selective extinction $R_\lambda=A_\lambda/E(B-V)$ for each band. Our results are 0.892, 0.553 and 0.363 for $J$, $H$ and $K_s$ respectively. As our systematic uncertainty of extinction corrections we adopt 0.02mag in $J$ and $H$ and 0.01mag in $K_s$ passband and this values should contain both uncertainty of the determined $E(B-V)$ values and the adopted reddening law for calculating $R_\lambda$.

Moreover, we calculated the quasi-magnitude Wesenheit index $W_{JKs} = K_s - R_{JK_s}\times(J-K_s)$ which, if the correct reddening law is used, should be independent of the reddening \citep{1982ApJ...253..575M}. We calculated $R_{JKs} = A_{Ks}/(A_{J}-A_{K_s})$ again using \citet{1989ApJ...345..245C} with the assumption of $R_{V} = 3.1$ which gives us $W_{JKs} = K_s - 0.69\times(J-K_s)$. The Wesenheit index should be in principle reddening free, however, there might be some systematic error related to the adopted reddening law. We assume this error to be at the level of 0.01mag.

\subsubsection{Distances}

The distances to our T2Ceps comes from $Gaia$ EDR3 parallaxes \citep[][column 3 of Table \ref{tab:t2cep_data}]{2021A&A...649A...2L} corrected for the zero-point offset (ZPO) as determined with dedicated \texttt{Python} code \footnote{\url{https://gitlab.com/icc-ub/public/gaiadr3_zeropoint}} described in \citet[][column 5 of Table \ref{tab:t2cep_data}]{2021A&A...649A...4L}. We increase the parallax uncertainties by 10$\%$ as suggested by \citet{2021ApJ...908L...6R} to account for possible excess uncertainty and resulting uncertainties are given in the column 4 of Table \ref{tab:t2cep_data}. According to \citet{2021A&A...649A...4L} the uncertainty of the zero-point is 5 $\mu as$ which is equivalent to 0.024mag in absolute magnitudes calculated using $Gaia$ EDR3 parallaxes for the median parallax of our sample (0.45 mas), so we adopt 0.024mag as a systematic uncertainty related to parallaxes zero-point on our absolute magnitudes and, by extension, on the PLRs determination. We adopt the Renormalised Unit Weight Error (RUWE) and the Goodness Of Fit (GOF) parameters as parallax quality indicators. These parameters were found to be relevant for CCeps by \citet{2021ApJ...913...38B} and \citet{2021ApJ...908L...6R}. RUWE should be close to 1, and not higher than 1.4 \citep{2021A&A...649A...2L} while good values of GOF according to \citet{2021ApJ...908L...6R} are below 12.5. Higher values of these parameters indicate e.g., saturation of the star in $Gaia$ photometry or its photocenter movement due to binarity. RUWE and GOF values for our T2Ceps are given in columns 6 and 7 of Table \ref{tab:t2cep_data}. Stars with RUWE $>$ 1.4 and GOF $>$ 12.5, namely TX Del, $\kappa$ Pav and ST Pup, are not used for PLRs calibration. Inverting parallaxes yields distances $d$ in kiloparsecs and from such distances we calculated distance moduli $\mu=5\log d -2$ (column 8 of Table \ref{tab:t2cep_data}). Distance moduli are subtracted from mean magnitudes (corrected for extinction) to obtain absolute magnitudes. We do not use these values directly to fit PLRs but we use Astrometric Based Luminosity instead, which is explained in the section \ref{subsec:plr}. Colors of markers in Fig. \ref{fig:mw_map} denote distances of studied T2Ceps.

\begin{deluxetable*}{cccccc}[b]
\tablenum{1}
\tablecaption{Near infrared photometry of Milky Way T2Ceps. Full version of this table is available as supplementary material.\label{tab:t2cep_phot}}
\tablewidth{0pt}
\tablehead{
\colhead{$Star$} & \colhead{$filter$} & \colhead{$HJD$} & \colhead{$m$} & \colhead{$\sigma_m$} & \colhead{source} \\
\colhead{} & \colhead{} & \colhead{} & \colhead{$(mag)$} & \colhead{$(mag)$} & \colhead{}
}
\startdata
AL Vir & J & 2458180.84288 & 8.486 & 0.012 & IRIS\\
AL Vir & J & 2458181.85912 & 8.477 & 0.012 & IRIS\\
AL Vir & J & 2458182.83384 & 8.379 & 0.018 & IRIS\\
AL Vir & J & 2458199.75768 & 8.370 & 0.030 & IRIS\\
AL Vir & J & 2458201.81562 & 8.486 & 0.016 & IRIS\\
AL Vir & J & 2458202.79249 & 8.458 & 0.019 & IRIS\\
AL Vir & J & 2458203.82118 & 8.324 & 0.011 & IRIS\\
AL Vir & J & 2458204.76877 & 8.176 & 0.013 & IRIS\\
AL Vir & J & 2458511.86068 & 8.359 & 0.015 & IRIS\\
AL Vir & J & 2458555.87695 & 8.115 & 0.010 & IRIS\\
AL Vir & J & 2458558.80783 & 8.208 & 0.030 & IRIS\\
AL Vir & J & 2458559.82360 & 8.300 & 0.017 & IRIS\\
AL Vir & J & 2458560.86504 & 8.353 & 0.030 & IRIS\\
AL Vir & J & 2458561.84425 & 8.462 & 0.030 & IRIS\\
AL Vir & J & 2458564.85894 & 8.174 & 0.015 & IRIS\\
AL Vir & J & 2458567.77684 & 8.128 & 0.030 & IRIS\\
AL Vir & J & 2458568.76634 & 8.215 & 0.013 & IRIS\\
AL Vir & J & 2458569.83901 & 8.288 & 0.030 & IRIS\\
AL Vir & J & 2458571.84169 & 8.428 & 0.030 & IRIS\\
AL Vir & J & 2458580.80998 & 8.342 & 0.030 & IRIS\\
AL Vir & J & 2458581.81409 & 8.435 & 0.014 & IRIS\\
AL Vir & J & 2458582.82497 & 8.469 & 0.014 & IRIS\\
AL Vir & J & 2458583.81693 & 8.470 & 0.030 & IRIS\\
AL Vir & J & 2458585.79749 & 8.146 & 0.014 & IRIS\\
AL Vir & J & 2458586.81269 & 8.089 & 0.030 & IRIS\\
AL Vir & J & 2458587.75078 & 8.115 & 0.030 & IRIS\\
AL Vir & J & 2458588.78209 & 8.144 & 0.011 & IRIS\\
AL Vir & J & 2458599.78980 & 8.170 & 0.012 & IRIS\\
AL Vir & H & 2458180.84864 & 8.084 & 0.013 & IRIS\\
AL Vir & H & 2458182.82839 & 8.228 & 0.030 & IRIS\\
AL Vir & H & 2458201.82129 & 8.101 & 0.030 & IRIS\\
AL Vir & H & 2458202.79827 & 8.132 & 0.010 & IRIS\\
AL Vir & H & 2458203.82692 & 8.191 & 0.016 & IRIS\\
AL Vir & H & 2458204.77439 & 8.054 & 0.015 & IRIS\\
AL Vir & H & 2458511.86675 & 8.203 & 0.016 & IRIS\\
AL Vir & H & 2458555.88273 & 7.860 & 0.030 & IRIS\\
AL Vir & H & 2458558.80238 & 7.879 & 0.015 & IRIS\\
AL Vir & H & 2458559.81815 & 7.920 & 0.019 & IRIS\\
AL Vir & H & 2458560.85966 & 8.014 & 0.030 & IRIS\\
AL Vir & H & 2458561.83892 & 8.118 & 0.012 & IRIS\\
AL Vir & H & 2458564.86471 & 7.931 & 0.012 & IRIS\\
AL Vir & H & 2458567.78254 & 7.855 & 0.030 & IRIS\\
\enddata
\end{deluxetable*}

\begin{longrotatetable}
\begin{deluxetable*}{ccccccccccccccccc}
\tablenum{2}
\tablecaption{Data of Milky Way T2Ceps\label{tab:t2cep_data}}
\tablewidth{0pt}
\tablehead{
\colhead{$Name$} & \colhead{$P$} & \colhead{$\pi$} & \colhead{$\sigma_{\pi}$} & \colhead{$ZPO$} &  \colhead{$RUWE$} & \colhead{$GOF$} & \colhead{$\mu$} & \colhead{$\sigma_{\mu}$} & \colhead{$<J>$} &  \colhead{$\sigma_J$} & \colhead{$<H>$} & \colhead{$\sigma_H$} & \colhead{$<K_{s}>$} & \colhead{$\sigma_{K_s}$} & \colhead{$E(B-V)$} & \colhead{$[Fe/H]$}\\
\colhead{} & \colhead{(days)} & \colhead{(mas)} & \colhead{(mas)} & \colhead{(mas)} &  \colhead{} & \colhead{} & \colhead{(mag)} & \colhead{(mag)} & \colhead{(mag)} &  \colhead{(mag)} & \colhead{(mag)} & \colhead{(mag)} & \colhead{(mag)} & \colhead{(mag)} & \colhead{(mag)} & \colhead{(dex)}
}
\decimalcolnumbers
\startdata
BX Del & 1.09180 & 0.3650 & 0.0150 & -0.0188 & 1.13 & 3.20 & 12.079 & 0.085 & 11.156 & 0.024 & 10.930 & 0.033 & 10.873 & 0.024 & 0.100 & -0.2\\
VY Pyx & 1.23995 & 3.9495 & 0.0186 & -0.0237 & 0.89 & -3.57 & 7.004 & 0.010 & 6.029 & 0.013 & 5.761 & 0.017 & 5.696 & 0.039 & 0.048 & -0.4\\
BL Her & 1.30744 & 0.8469 & 0.0179 & 0.0016 & 1.29 & 10.20 & 10.365 & 0.046 & 9.206 & 0.014 & 9.017 & 0.015 & 8.932 & 0.013 & 0.067 & -0.1\\
KZ Cen & 1.52004 & 0.3024 & 0.0153 & -0.0224 & 1.20 & 5.51 & 12.442 & 0.103 & 11.302 & 0.033 & 11.051 & 0.027 & 10.987 & 0.030 & 0.084 & -\\
SW Tau & 1.58355 & 1.2244 & 0.0222 & -0.0103 & 1.25 & 5.52 & 9.542 & 0.039 & 8.320 & 0.007 & 8.082 & 0.007 & 7.955 & 0.008 & 0.252 & 0.2\\
V971 Aql & 1.62453 & 0.4400 & 0.0219 & -0.0175 & 1.26 & 5.10 & 11.698 & 0.105 & 10.508 & 0.014 & 10.199 & 0.016 & 10.112 & 0.017 & 0.174 & -\\
DU Ara & 1.63949 & 0.3394 & 0.0180 & -0.0189 & 1.25 & 7.52 & 12.228 & 0.110 & 10.923 & 0.017 & 10.665 & 0.015 & 10.580 & 0.024 & 0.055 & -\\
V439 Oph & 1.89298 & 0.4753 & 0.0163 & -0.0103 & 1.17 & 4.68 & 11.569 & 0.073 & 10.367 & 0.014 & 9.969 & 0.016 & 9.855 & 0.020 & 0.268 & -\\
RT TrA & 1.94612 & 1.0162 & 0.0162 & -0.0021 & 0.95 & -1.85 & 9.961 & 0.035 & 8.553 & 0.014 & 8.331 & 0.019 & 8.200 & 0.015 & 0.112 & - \\
V553 Cen & 2.06055 & 1.7286 & 0.0224 & -0.0080 & 0.94 & -0.97 & 8.801 & 0.028 & 7.242 & 0.016 & 6.976 & 0.017 & 6.869 & 0.017 & 0.069 & -\\
AU Peg & 2.41174 & 1.6463 & 0.0200 & -0.0180 & 1.25 & 6.79 & 8.894 & 0.026 & 7.787 & 0.014 & - & - & 7.066 & 0.015 & 0.046 & -0.2\\
FM Del & 3.95373 & 0.2300 & 0.0135 & -0.0174 & 0.93 & -1.71 & 13.033 & 0.119 & 11.08 & 0.017 & 10.729 & 0.024 & 10.646 & 0.017 & 0.087 & -\\
TX Del & 6.16742 & 0.9078 & 0.0294 & -0.0112 & 1.94 & 27.9 & 10.183 & 0.070 & 7.836 & 0.011 & - & - & 7.460 & 0.011 & 0.085 & 0.5\\
$\kappa$ Pav & 9.07890 & 5.2451 & 0.1221 & 0.0046 & 2.29 & 37.6 & 6.403 & 0.051 & 3.201 & 0.021 & 2.881 & 0.021 & 2.784 & 0.021 & 0.019 & -\\
AL Vir & 10.30611 & 0.4574 & 0.0190 & -0.0152 & 0.98 & -0.32 & 11.627 & 0.088 & 8.270 & 0.011 & 7.956 & 0.013 & 7.887 & 0.013 & 0.072 & -0.4\\
AP Her & 10.38432 & 0.3578 & 0.0152 & -0.0028 & 1.13 & 3.74 & 12.215 & 0.092 & 9.034 & 0.032 & 8.630 & 0.030 & 8.520 & 0.030 & 0.373 & -0.7\\
CO Pup & 16.04266 & 0.3471 & 0.0166 & -0.0212 & 1.31 & 9.29 & 12.169 & 0.115 & 8.975 & 0.017 & 8.534 & 0.013 & 8.415 & 0.019 & 0.155 & -0.6\\
W Vir & 17.27137 & 0.4728 & 0.0222 & -0.0227 & 1.06 & 2.13 & 11.525 & 0.098 & 8.442 & 0.012 & 8.117 & 0.014 & 8.029 & 0.016 & 0.036 & -1.0\\
ST Pup & 18.7010 & 0.4099 & 0.0232 & -0.0091 & 2.07 & 25.6 & 11.889 & 0.121 & 8.472 & 0.008 & 8.118 & 0.007 & 7.944 & 0.011 & 0.120 & -\\
MR Ara & 19.81649 & 0.2087 & 0.0236 & -0.0080 & 0.97 & -0.75 & 13.321 & 0.237 & 9.817 & 0.032 & 9.421 & 0.059 & 9.279 & 0.041 & 0.110 & -\\
RS Pav & 19.96997 & 0.3547 & 0.0157 & -0.0003 & 0.99 & -0.26 & 12.249 & 0.097 & 8.799 & 0.018 & 8.353 & 0.02 & 8.231 & 0.018 & 0.072 & -\\
\enddata
\end{deluxetable*}
\end{longrotatetable}

\begin{widetext}

\begin{figure}[b!]
\plotone{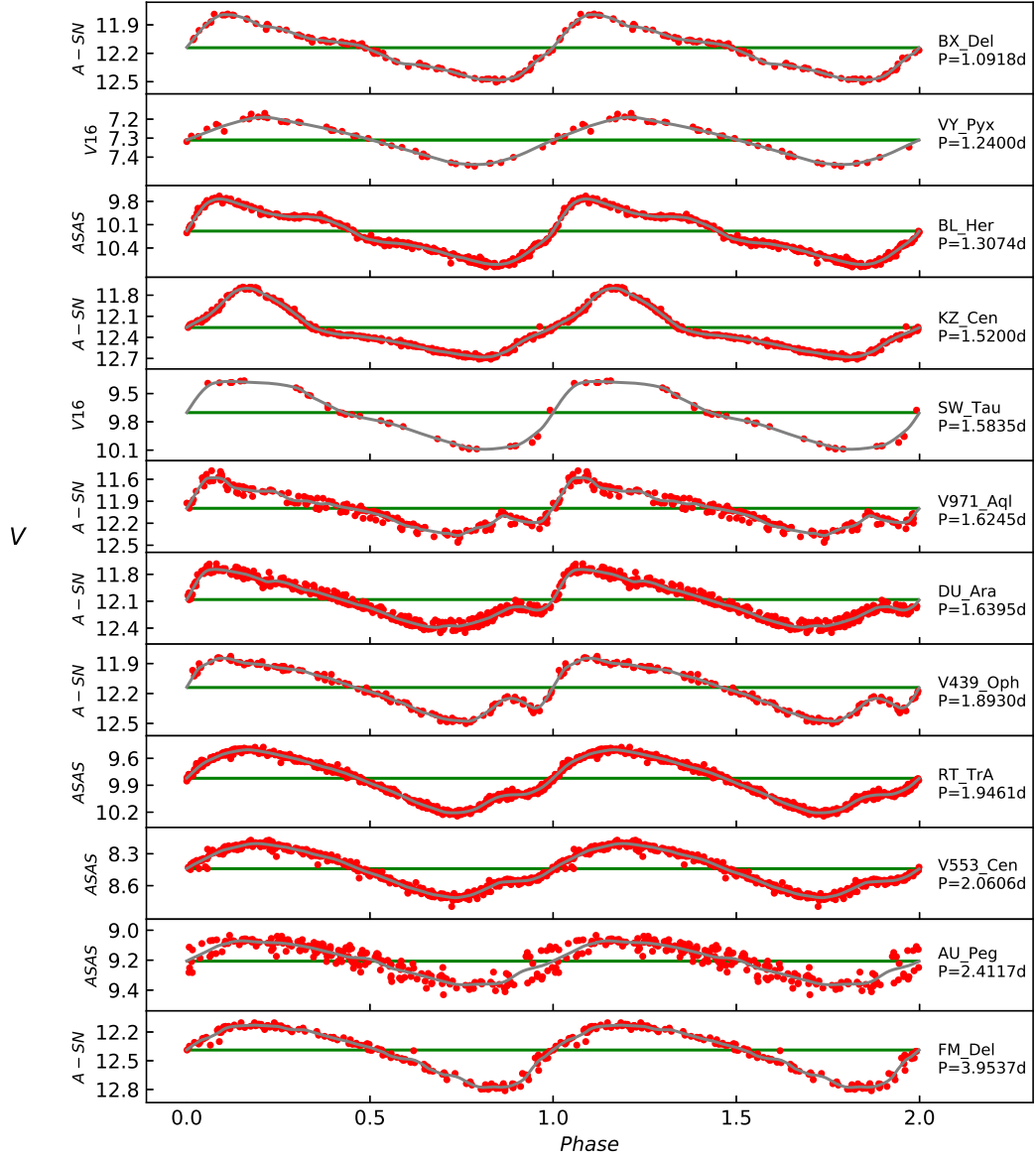}
\caption{$V$-band light curves of BL Her stars. Meaning of labels is as follows: A-SN corresponds to ASAS-SN data, V16 denotes VYSOS 16 and B08 is Berdnikov (2008). Green line is the mean magnitude.\label{fig:blher_vlc}}
\end{figure}

\begin{figure}
\plotone{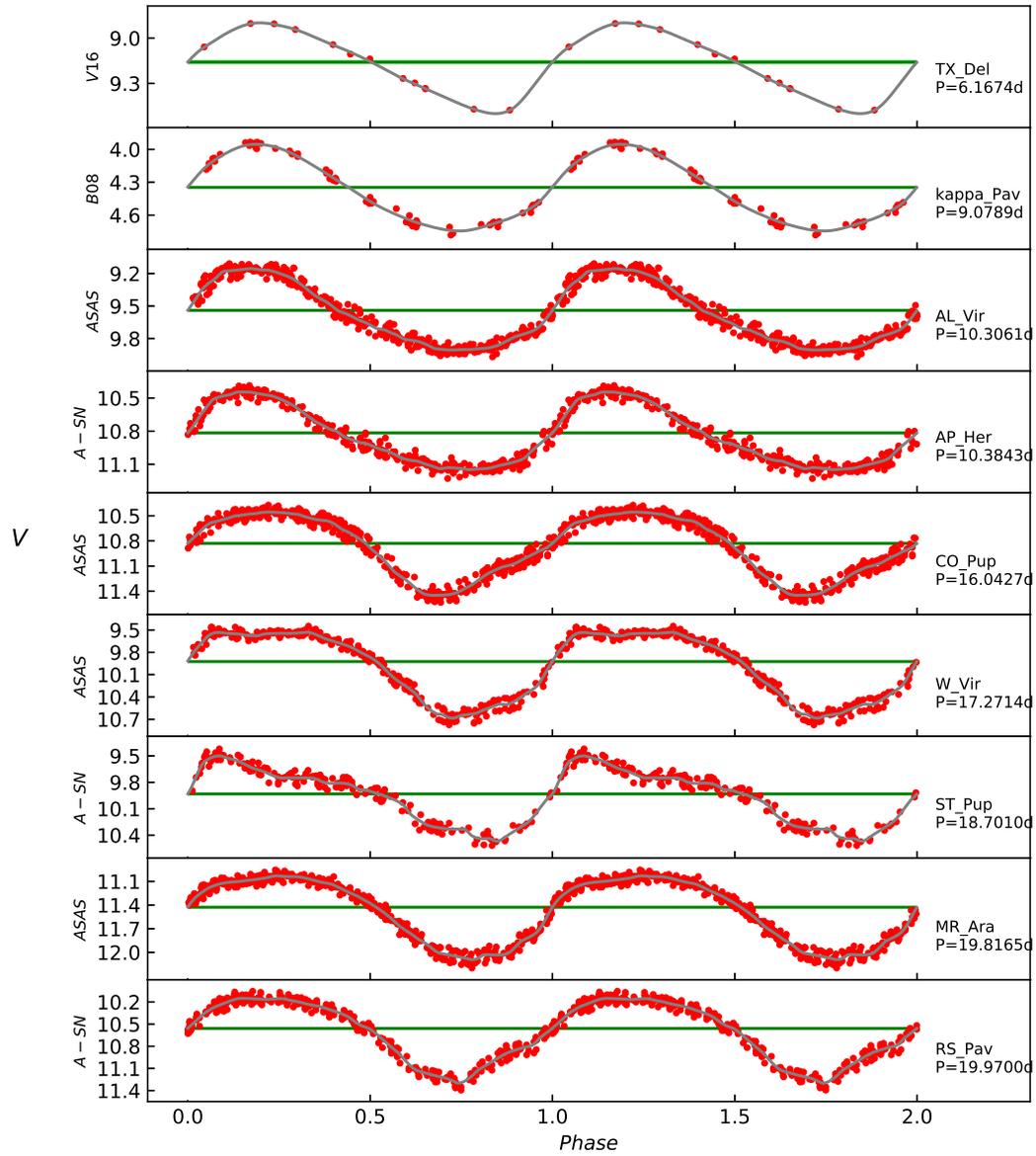}
\caption{$V$-band light curves of W Vir stars. Meaning of labels is the same as in Fig. \ref{fig:blher_vlc}.\label{fig:wvir_vlc}}
\end{figure}

\begin{figure}
\plotone{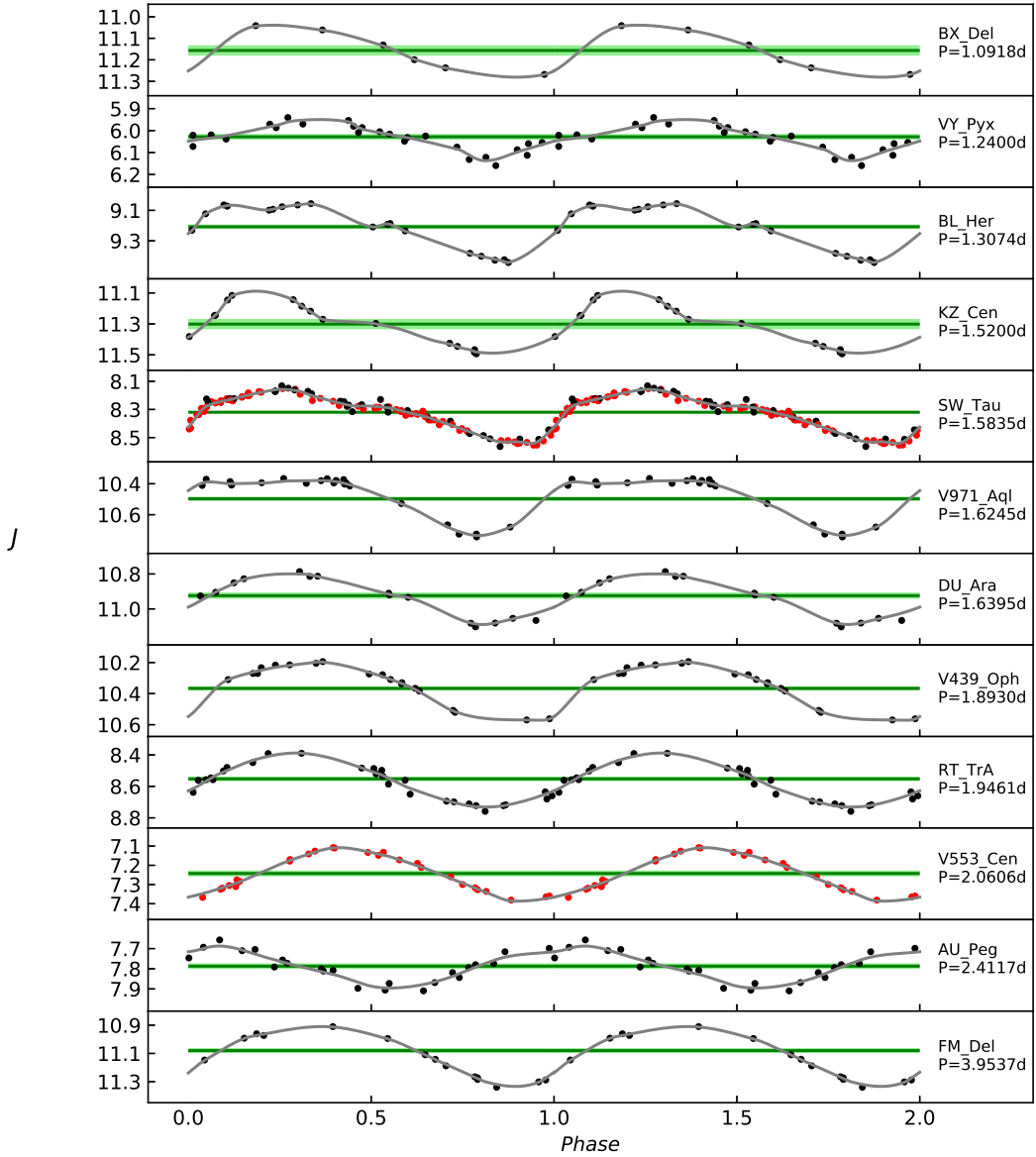}
\caption{$J$-band light curves of BL Her stars. Black points are measurements from IRIS, red are from \citet{2008MNRAS.386.2115F}. Green line is the mean magnitude while light-green denotes its $1\sigma$ uncertainty. \label{fig:blher_jlc}}
\end{figure}

\begin{figure}
\plotone{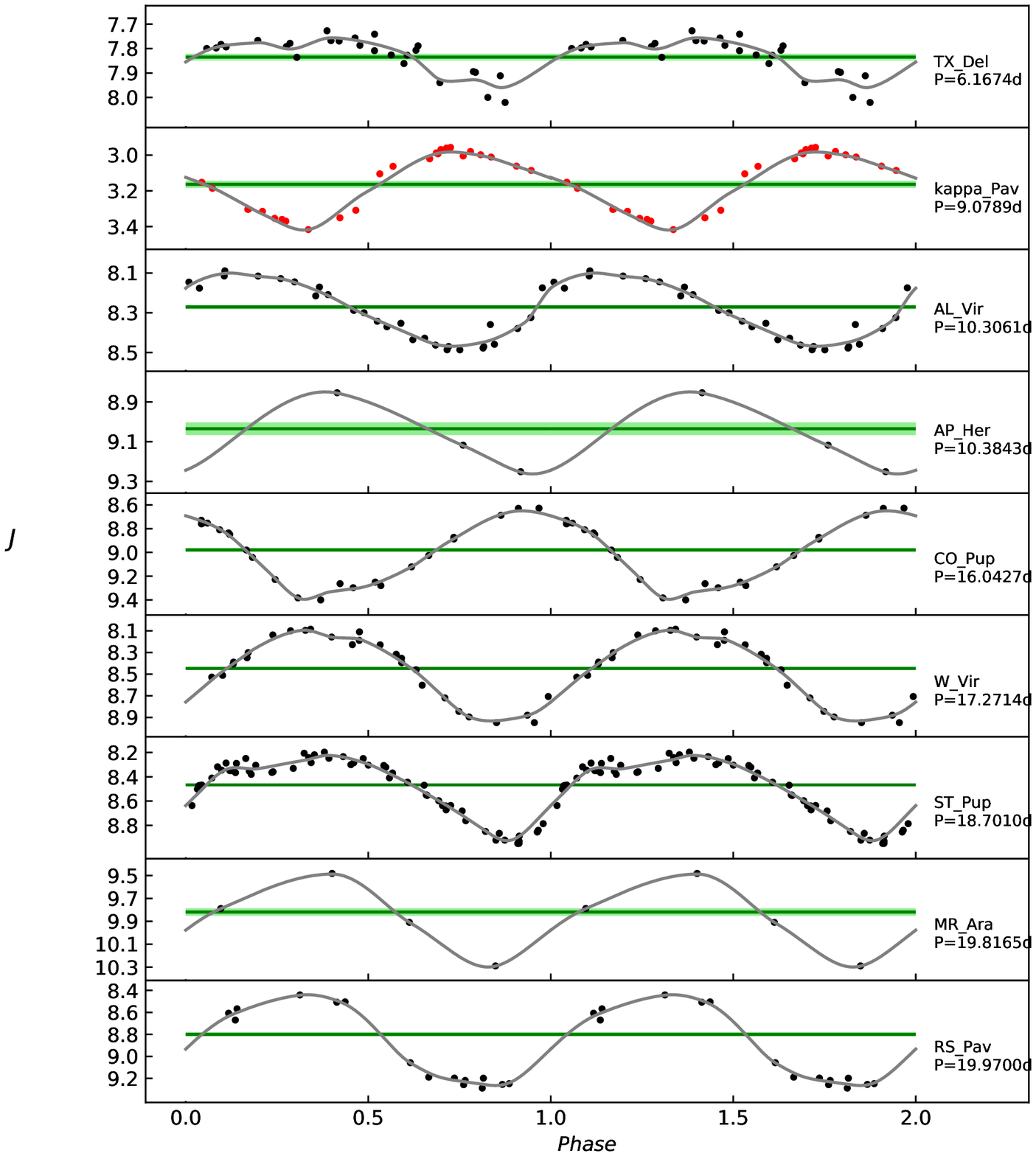}
\caption{$J$-band light curves of W Vir stars. Meaning of colors and lines is the same as in Fig. \ref{fig:blher_jlc}. \label{fig:wvir_jlc}}
\end{figure}

\begin{figure}
\plotone{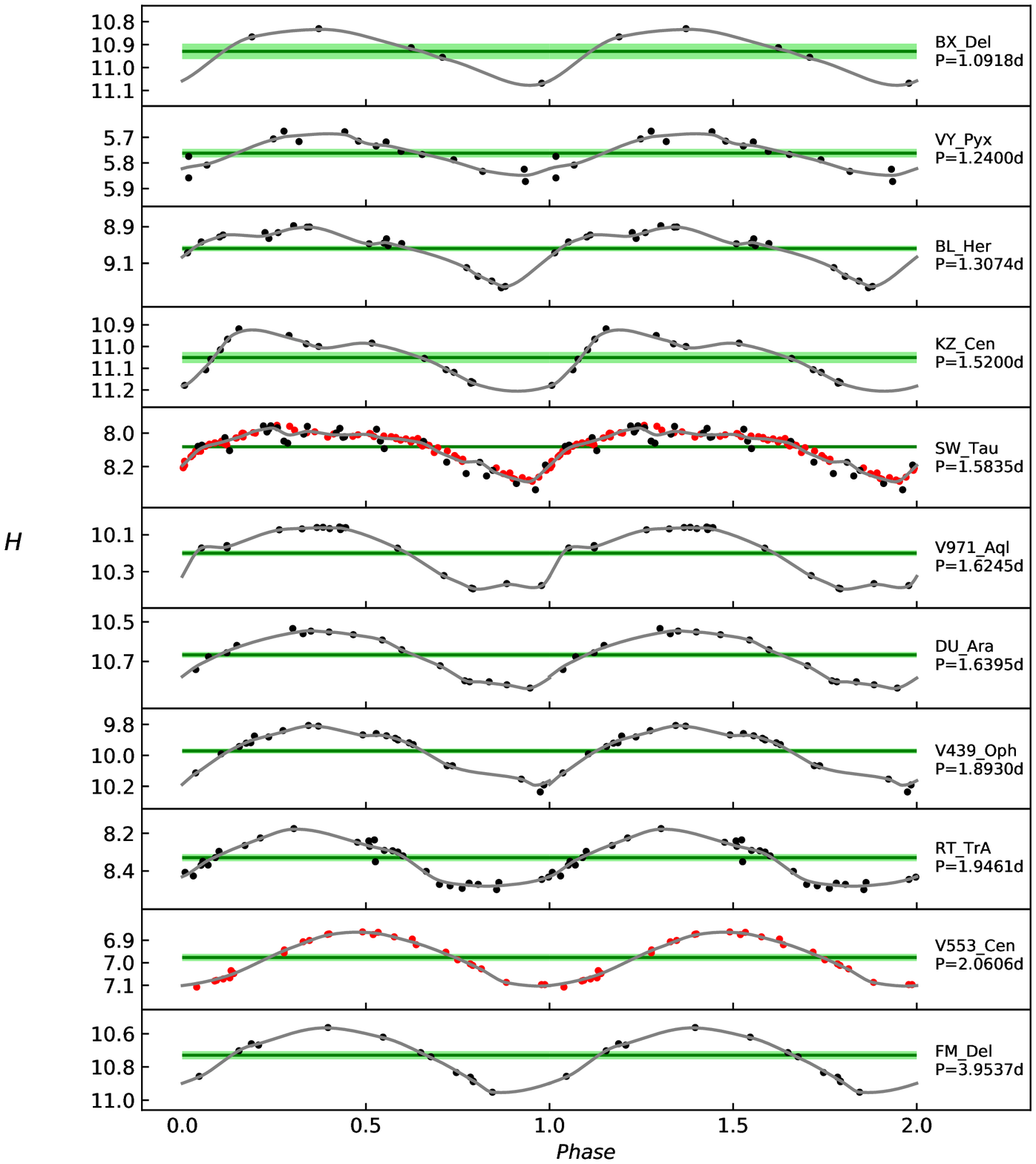}
\caption{$H$-band light curves of BL Her stars. Meaning of colors and lines is the same as in Fig. \ref{fig:blher_jlc} \label{fig:blher_hlc}}
\end{figure}

\begin{figure}
\plotone{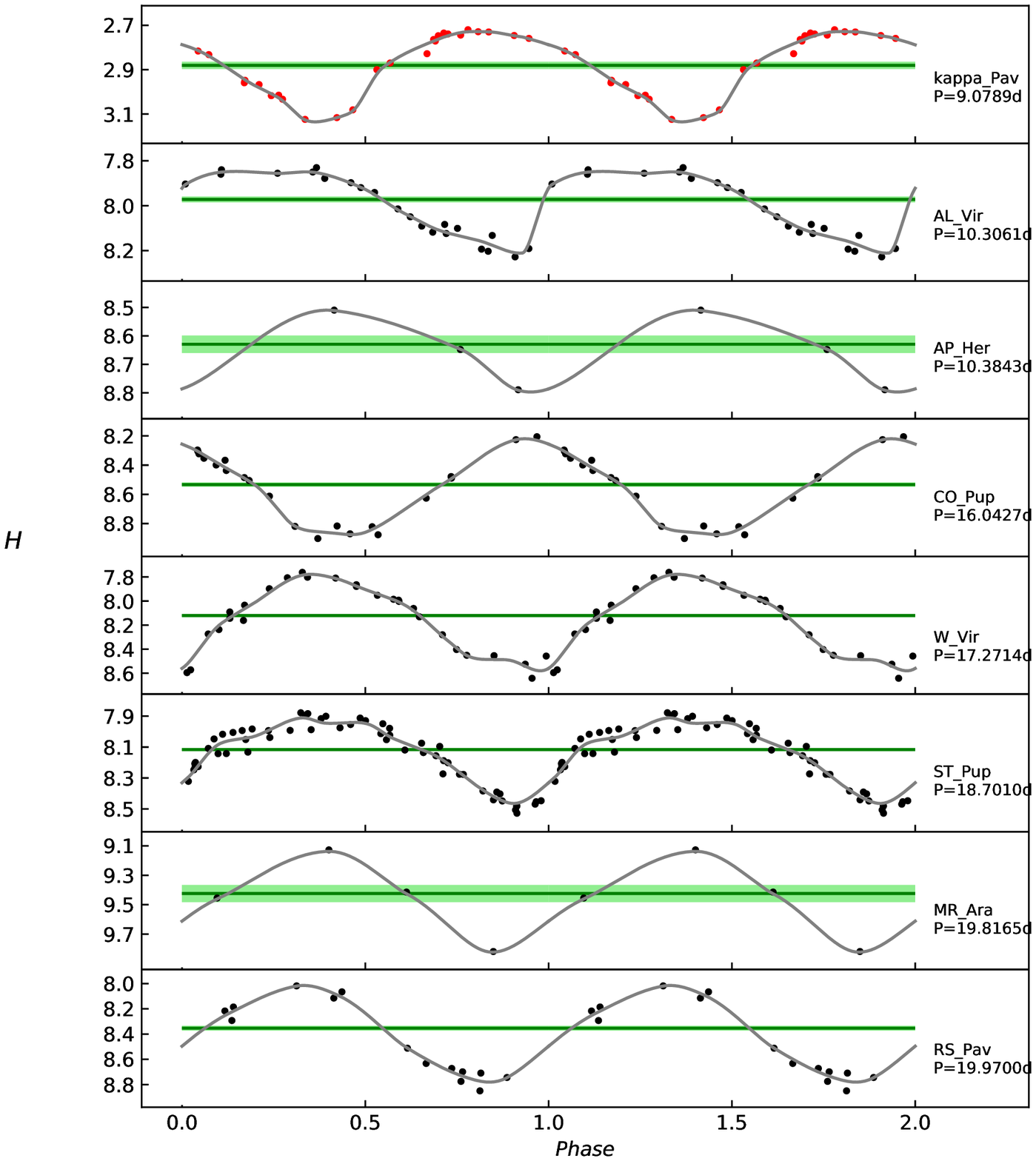}
\caption{$H$-band light curves of W Vir stars. Meaning of colors and lines is the same as in Fig. \ref{fig:blher_jlc} \label{fig:wvir_hlc}}
\end{figure}

\begin{figure}
\plotone{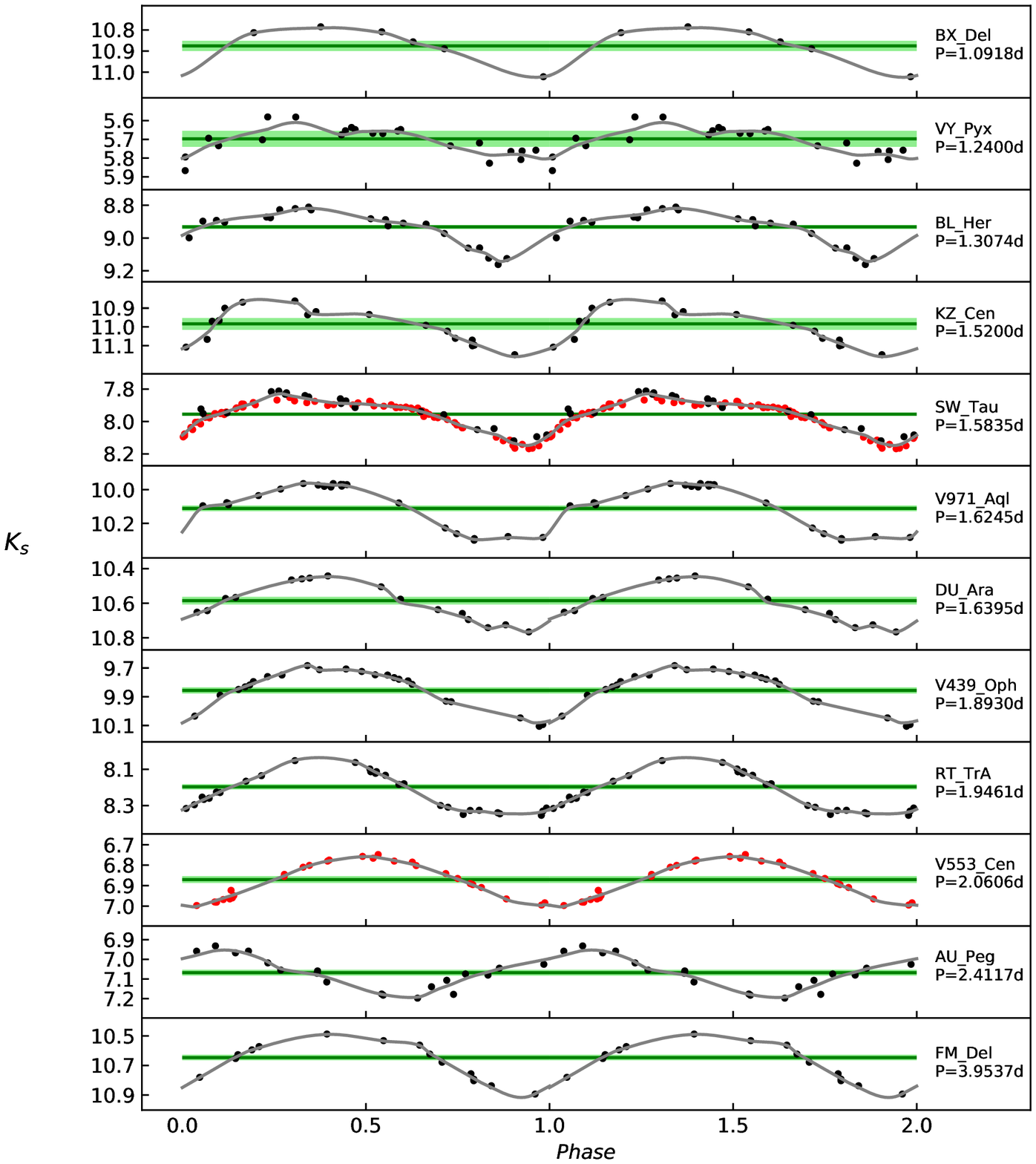}
\caption{$K_s$-band light curves of BL Her stars. Meaning of colors and lines is the same as in Fig. \ref{fig:blher_jlc} \label{fig:blher_klc}}
\end{figure}

\begin{figure}
\plotone{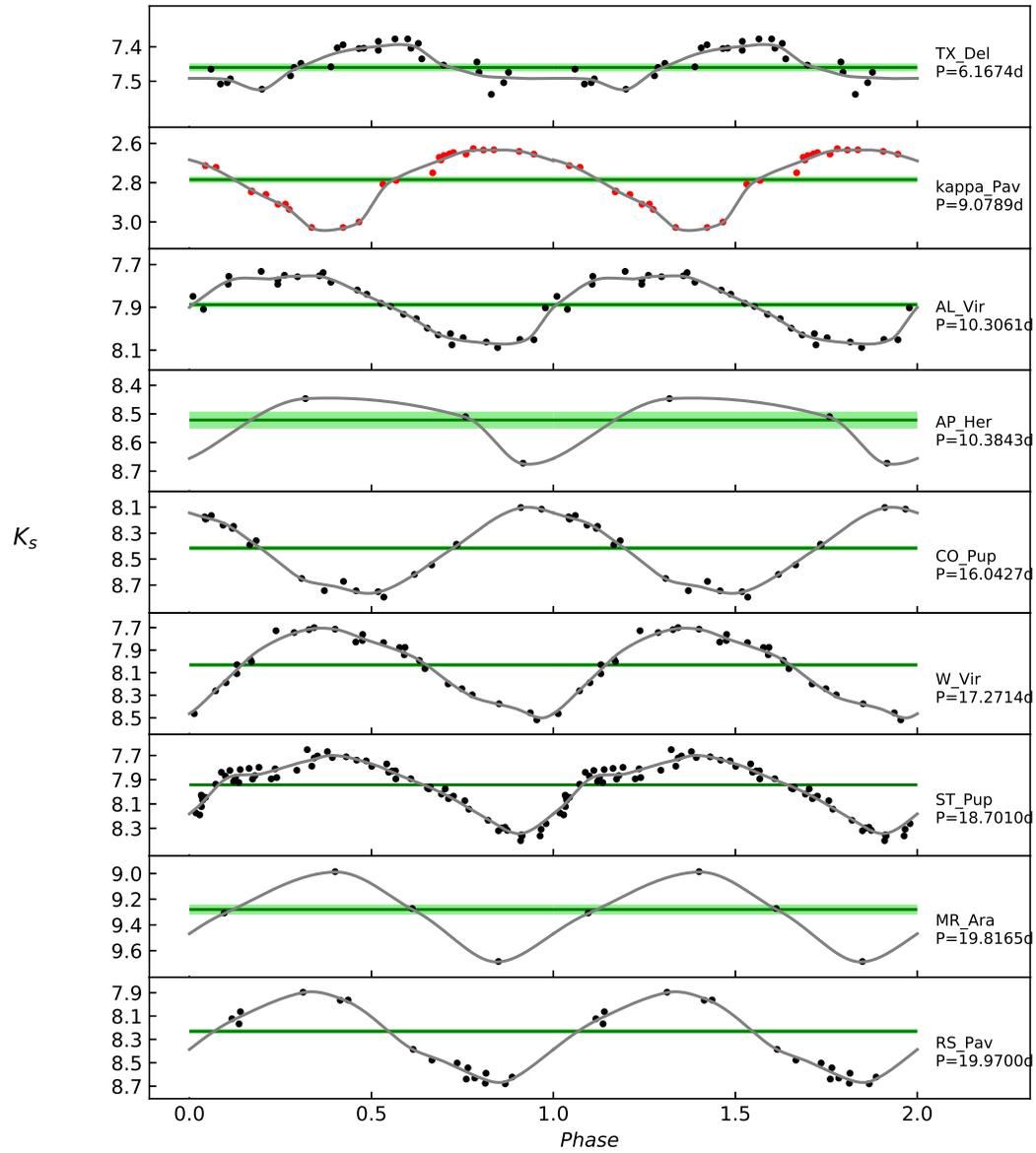}
\caption{$K_s$-band light curves of W Vir stars. Meaning of colors and lines is the same as in Fig. \ref{fig:blher_jlc} \label{fig:wvir_klc}}
\end{figure}

\end{widetext}

\clearpage
\subsection{The Large Magellanic Cloud} \label{subsec:mcdata}

\subsubsection{Sample selection, periods and photometry}

\begin{figure}[htb]
\epsscale{1.2}
\plotone{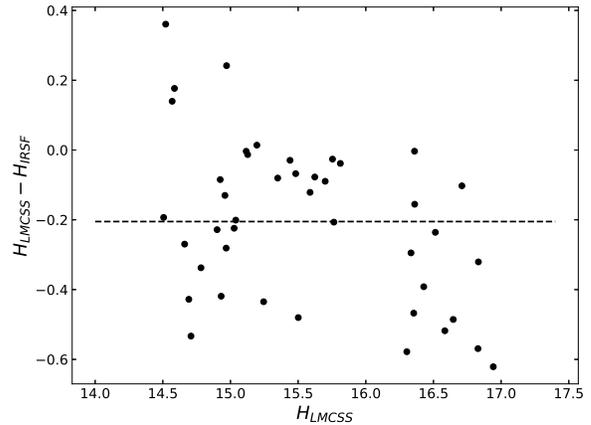}
\caption{Comparison of LMC T2Ceps H-band mean magnitudes from LMCSS and IRSF catalogues. Dashed line represents the mean difference between these two sources of about -0.2mag.\label{fig:lmcss_irsf_comp}}
\end{figure}

The most complete list of T2Ceps in the LMC is the OGLE catalogue \citep{2018AcA....68...89S} so we adopt classification and periods from this source (we exclude RV Tau and pW Vir stars from this initial sample). Infrared photometry from several sources is available e.g. Vista Magellanic Clouds Survey \citep[VMC][]{2015MNRAS.446.3034R}, Large Magellanic Cloud Synoptic Survey \citep[LMCSS][]{2017AJ....153..154B} and the InfraRed Survey Facility \citep[IRSF][]{2011MNRAS.413..223M}. We preferentially use mean magnitudes determined from time-series photometry over single-point measurement, so we use VMC for $J$ and $K_s$ passbands ($H$ band is not provided). We adopt mean magnitudes of T2Ceps from Table 4 in \citet{2015MNRAS.446.3034R}. For $H$ band we initially used LMCSS data but obtained a LMC distance modulus that was significantly smaller ($\sim$0.2mag) than in $J$ and $K_s$ passbands while IRSF phase-corrected $H$-band magnitudes give a value consistent with $J$ and $K_s$. In Fig. \ref{fig:lmcss_irsf_comp} we plot the difference between H-band LMCSS mean magnitudes \citep[data from Table 4 in ][]{2017AJ....153..154B} and IRSF \citep[data from Table 1 in ][]{2011MNRAS.413..223M} phase-corrected magnitudes for T2Ceps present in both sources and the mean difference between the two samples is close to -0.2mag. We do not know the origin of this discrepancy and we present results using both LMCSS and IRSF photometry, but the IRSF sample is used for further analysis. Uncertainties of IRSF magnitudes given in \citet{2011MNRAS.413..223M} in their Table 1 do not take into account statistical uncertainties related to the phase-correction. We assume that they are at the level of 0.05mag and we add this value quadratically to the given $H$-band uncertainty.

Photometry of MW T2Ceps is in the 2MASS system, thus we have to transform LMC data into 2MASS as well. LMCSS photometry is already provided in the 2MASS system. IRSF was transformed using formulae given in \citet{2007PASJ...59..615K}. VMC photometry is in the VISTA system so we transformed it to the 2MASS using respective formulae from CASU\footnote{\url{http://casu.ast.cam.ac.uk/surveys-projects/vista/}}. We adopt 0.02mag as a systematic uncertainty for the VMC (photometric zero-point accuracy) mean magnitudes. \citet{2015MNRAS.446.3034R} found a difference of about 0.05mag between T2Ceps PLRs zero-points in $J$ and $K_s$ based on the VMC mean magnitudes and the IRSF phase-corrected data. The difference is caused most probably mainly by the systematic error associated with applying phase-correction for IRSF single-point observation based on $I$-band light curve amplitude. We then adopt 0.05mag as a systematic uncertainty for the IRSF $H$-band mean magnitudes.

\begin{figure}[htb]
\epsscale{1.2}
\plotone{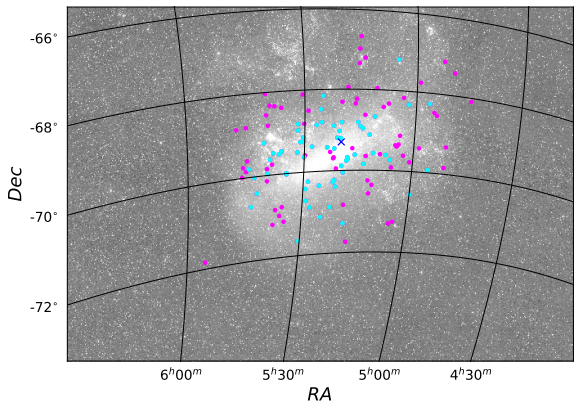}
\caption{Positions of T2Ceps in the Large Magellanic Cloud considered in this study. Cyan points mark stars observed by the IRSF solely, while magenta points are stars present both in IRSF and VMC catalogues.\label{fig:lmc_map}}
\end{figure}

Fig. \ref{fig:lmc_map} presents a map of the LMC, with marked positions of our final sample of T2Ceps and Table \ref{tab:lmc_t2cep_data} contains compilation of data used in the analysis.

\subsubsection{Extinction and distances}

We estimate extinction towards LMC T2Ceps using recent maps of \citet{2020ApJ...889..179G} based on Red Clump stars. From the map we obtain $E(B-V)$ color excess for each star (column 11 of Table \ref{tab:lmc_t2cep_data}). We use again \citet{1989ApJ...345..245C} and \citet{1994ApJ...422..158O} and $R_V$=3.1 to calculate total extinction in each photometric band. Statistical errors of $E(B-V)$ given in the map are below 0.01mag, but we adopt 0.01mag for each star. Systematic error of the reddening is related mostly to the uncertainty of the intrinsic color of the Red Clump. \citet{2020ApJ...889..179G} estimated the uncertainty of the intrinsic Red Clump color in the LMC to be 0.013mag for $(B-V)$ color index, thus we adopt 0.01mag for the systematic uncertainty related to the reddening in each band.  

The most accurate distance to the LMC was obtained using late-type eclipsing binaries by \citet{2019Natur.567..200P}. They obtained the distance modulus of 18.477$\pm$0.004(stat.)$\pm$0.026(syst.)mag and such value is subtracted from reddening-corrected mean magnitudes in order to calculate absolute magnitudes. We also calculated individual corrections related to the geometry of the LMC. We adopt the plane-disc model of the LMC from \citet{2001AJ....122.1807V}. We assume the same definition of the LMC center as in \citet{2019Natur.567..200P} ($\alpha_0$=5$^h$20$^m$, $\delta_0$=-69$^{\circ}$18$'$), the LMC inclination $i$=25$^{\circ}$ and the line-of-nodes position angle $\theta$=132$^{\circ}$ \citep[also from][]{2019Natur.567..200P}. Calculated corrections are presented in column 12 of Table \ref{tab:lmc_t2cep_data} and they are subtracted from absolute magnitudes. Distribution of RR Lyrae stars in the LMC \citep{2017AcA....67....1J,2021MNRAS.504....1C} show that the old population stars form a spherical halo around the LMC dics, thus the line-of-sight depth effect should have a significant impact on the spread of LMC T2Ceps PLRs, however, it is not possible to apply any corrections for this effect. We suspect that the depth effect is at the level of 0.1mag, so we add quadratically such value to the uncertainty of each star.

\begin{deluxetable*}{cccccccccccc}[htb!]
\tablenum{3}
\tablecaption{Data of T2Ceps in the LMC. Full version of this table is available as supplementary material.\label{tab:lmc_t2cep_data}}
\tablehead{
\colhead{$OGLE$ $id^1$} & \colhead{$P$} & \colhead{$<J>$} & \colhead{$\sigma_J$} & \colhead{$<H_{LMCSS}>$} & \colhead{$\sigma_{HLMCSS}$} & \colhead{$<H_{IRSF}>$} & \colhead{$\sigma_{HIRSF}$} & \colhead{$<K_s>$} & \colhead{$\sigma_{Ks}$} & \colhead{$E(B-V)$} & \colhead{$\Delta m_{geom}$}\\
\colhead{} & \colhead{(days)} & \colhead{(mag)} & \colhead{(mag)} & \colhead{(mag)} & \colhead{(mag)} & \colhead{(mag)} & \colhead{(mag)} & \colhead{(mag)} & \colhead{(mag)} & \colhead{(mag)} & \colhead{(mag)}
\\
\colhead{(1)} & \colhead{(2)} & \colhead{(3)} & \colhead{(4)} & \colhead{(5)} & \colhead{(6)} & \colhead{(7)} & \colhead{(8)} & \colhead{(9)} & \colhead{(10)} & \colhead{(11)} & \colhead{(12)}
}
\startdata
053 & 1.0429993 & - & - & - & - & 17.299 & 0.05 & - & - & 0.131 & 0.003\\
188 & 1.0493152 & - & - & - & - & 17.576 & 0.09 & - & - & 0.179 & -0.025\\
006 & 1.0879277 & - & - & - & - & 17.450 & 0.07 & - & - & 0.105 & 0.034\\
020 & 1.1081258 & - & - & - & - & 17.111 & 0.06 & - & - & 0.104 & 0.020\\
071 & 1.1521741 & 17.536 & 0.022 & 16.832 & 0.151 & 17.297 & 0.05 & 17.324 & 0.026 & 0.117 & 0.004\\
089 & 1.1673093 & 17.732 & 0.018 & 16.829 & 0.050 & 17.209 & 0.09 & 17.476 & 0.043 & 0.104 & 0.006\\
\enddata
\tablecomments{$^1$ full OGLE id contains prefix OGLE-LMC-T2CEP}
\end{deluxetable*}

\section{Analysis and results} \label{sec:results}

\subsection{MW}\label{subsec:plr}

\begin{figure}[]
\epsscale{1.2}
\plotone{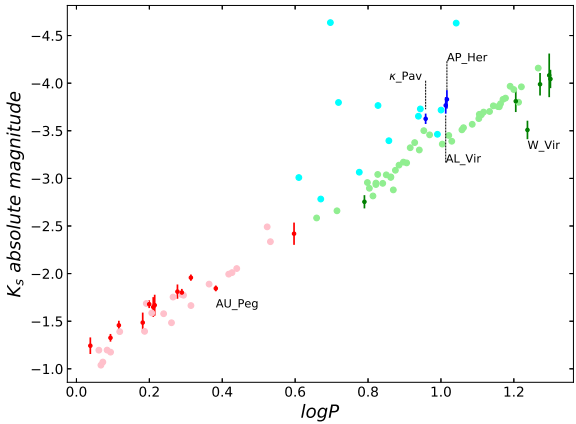}
\caption{$K_s$ band absolute magnitudes as a function of periods of pulsation for Milky Way and Large Magellanic Cloud T2Ceps. Red, green and blue datapoints with errorbars are Galactic BL Her, W Vir and peculiar W Vir stars respectively. Pink points are BL Her stars, light green W Vir stars and cyan denotes peculiar W Vir stars in the LMC \citep[source][]{2015MNRAS.446.3034R} shifted to obtain absolute magnitudes using LMC distance modulus from \citet{2019Natur.567..200P}.\label{fig:permag}}
\end{figure}

\begin{figure}[]
\epsscale{1.2}
\plotone{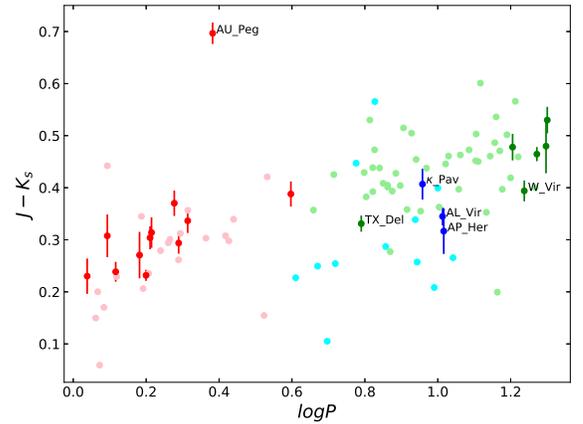}
\caption{Color index of the Milky Way and LMC T2Ceps as a function of pulsational period. Markers colors are the same as in Fig. \ref{fig:permag}.\label{fig:percol}}
\end{figure}

Before determining PLRs, we have to take a closer look at our MW sample to check if it contains pW Vir stars. As we mentioned in the introduction, pW Vir stars are usually more luminous and bluer than regular W Vir stars of similar periods, and have different light curve morphologies. Visually comparing the shapes of optical light curves of W Vir stars with LMC T2Ceps optical light curves from the OGLE Atlas of Variable Star Light Curves \footnote{\url{http://ogle.astrouw.edu.pl/atlas/}} \citep{2018AcA....68...89S} we suspect that three stars: $\kappa$ Pav, AP Her and AL Vir, are pW Vir stars. $\kappa$ Pav has already been suggested to be a member of pW Vir subgroup by \citet{2008MNRAS.386.2115F}. \citet{2015A&A...576A..64B} studied this star in detail using photometric, interferometric and spectroscopic data and did not find signatures of a companion star, while it is believed that pW Vir stars are binaries \citep{2017AcA....67..297S}. On the other hand, $RUWE$ and $GOF$ parallax quality parameters of $\kappa$ Pav are higher than the recommended values, which can be the result of its binarity. Fig. \ref{fig:permag} and \ref{fig:percol} present $K_s$ band absolute magnitude and $J-K_s$ color index as a function of period for our MW sample (red, green and blue denotes BL Her, W Vir and pW Vir stars respectively) and LMC BL Her (pink), W Vir (light green) and pW Vir (cyan) stars from the VMC survey \citep{2015MNRAS.446.3034R}. Absolute magnitudes of the LMC sample were calculated by correcting VMC mean magnitudes for extinction using \citet{2020ApJ...889..179G} reddening maps, the \citet{1989ApJ...345..245C} reddening law and the LMC distance modulus of 18.477mag from \citet{2019Natur.567..200P}. The three stars mentioned above are overlapping with a "clump" of pW Vir stars in the LMC in  Fig. \ref{fig:permag}. AL Vir and AP Her are also bluer than other W Vir stars in our sample, lying in a range of colors of the LMC pW Vir stars, while $\kappa$ Pav is rather located among W Vir, although some LMC pW Vir stars are also placed among W Vir stars on this diagram. This, in our opinion, confirms that AP Her and AL Vir are pW Vir type stars and $\kappa$ Pav is most probably peculiar as well so we do not consider these stars in further analysis. 

Another interesting object from the point of view of colors is also AU Peg; it is very red ($J-K_s\approx$0.7mag) compared to other BL Her type stars ($J-K_s\approx$0.3mag). \citet{1984AJ.....89..119H} found this star to be a member of a single-lined spectroscopic binary with a more massive compact object as a companion and a dusty envelope surrounding the system. Its optical light curve (Fig. \ref{fig:blher_vlc}) presents significant instabilities in pulsations and \citet{2007AN....328..837J} suggested that this star is a double mode CCep, but its position on our period-luminosity diagram shows clearly that it is a T2Cep.  Further analysis shows that deviation of AU Peg from the PLR decrease for longer wavelengths and vanishes in the Wesenheit index. It suggest that this deviation is caused by high extinction, much higher than estimated from \citet{2011ApJ...737..103S}. We decided to exclude this star from fitting PLRs in $J$, $H$ and $K_s$ passbands. According to \citet{1984AJ.....89..119H} the contribution of the non-pulsating component of this system to the total brightness is very low so we include AU Peg in our PLRs fit for the Wesenheit index (excluding AU Peg in the Wesenheit index do not change our results).

Another star which was found to be in a binary system is TX Del \citep{1989AJ.....98..981H}. This star is not included in the PLR calibration as its RUWE and GOF parameters are higher than acceptable values but it is interesting to discuss its observed properties. Its position on the period-luminosity diagram does not deviate from the relation determined by the LMC sample but it is on the blue side of the color index range of LMC W Vir stars, thus it is another pW Vir candidate.

W Virginis star- a subgroup prototype- is yet another star drawing attention. It is significantly fainter (0.2-0.5mag) than other W Vir type stars on the period-luminosity diagram. Its deviation from the PLR is higher for longer wavelengths and in the Wesenheit index. On the other hand, its position on the period-color diagram does not deviate significantly from other W Vir stars but it is located among bluest LMC W Vir stars. \citet{2011A&A...526A.116K} analysed emission lines in W Virginis spectrum and concluded that this star is surrounded by an envelope. Such an envelope could be a possible explanation for the faintness, however, the wavelength dependence of the effect is opposite to what is expected from extinction. This star is also exceptional as it shows the period doubling effect in its light curves \citep{2007AJ....134.1999T}, which makes it similar to RV Tau stars. We decided to determine PLRs including and excluding this star (case 1 and 2, respectively). We also calibrate PLRs for BL Her stars solely (case 3). 

\begin{widetext}

\begin{figure}[b!]
\epsscale{1.}
\plotone{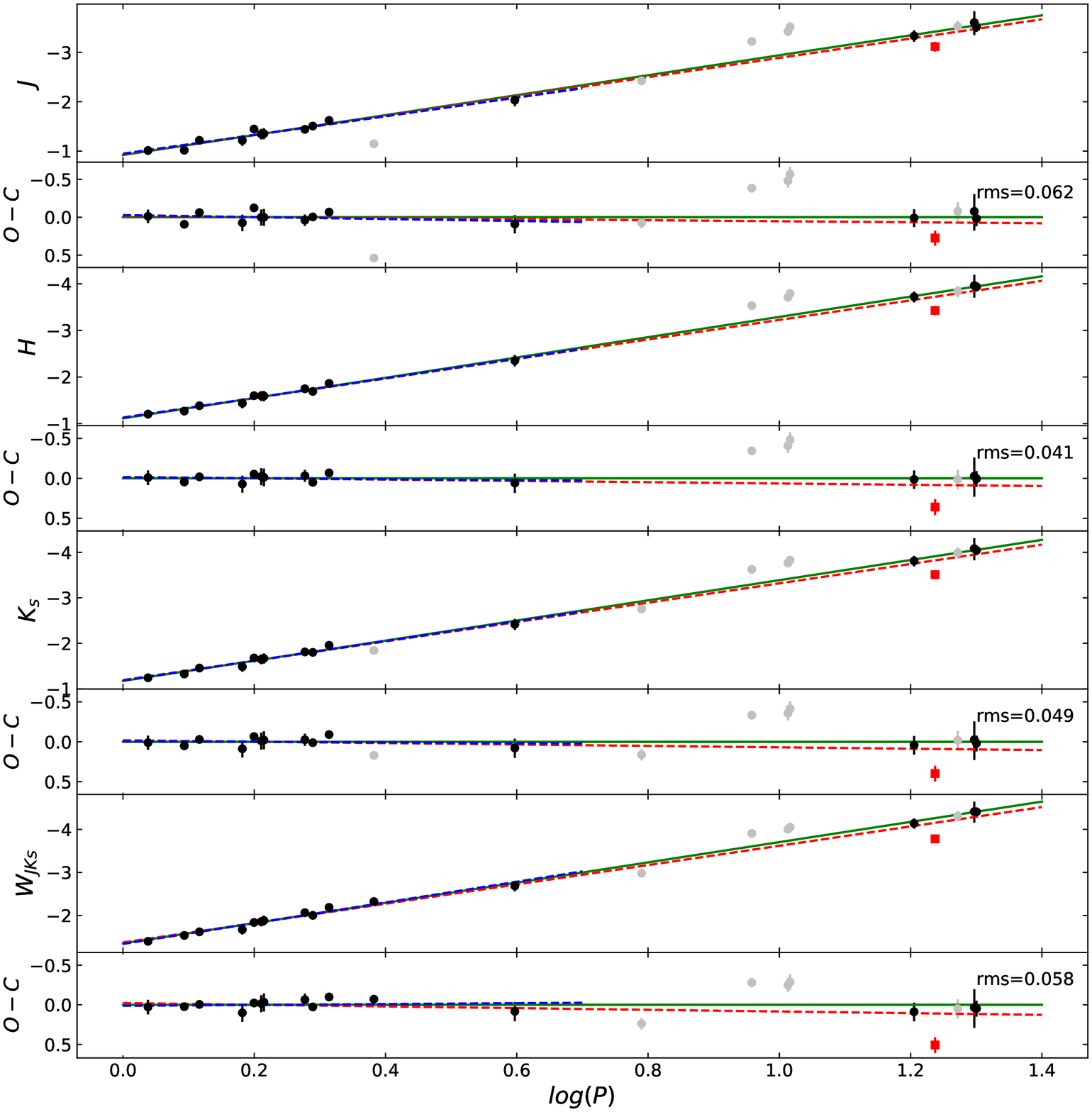}
\caption{Absolute mean magnitues of Milky Way T2Ceps in $J$, $H$ and $K_s$ and $W_{JKs}$ wesenheit index plotted against logarithm of periods. Black and grey points mark stars used and excluded from fitting procedure, respectively. The red square is W Virginis star and the red line represent fit including this star, while the green line is obtained excluding it. The blue line is obtained based on BL Her type stars solely. O-C plots present deviations from respective green lines.\label{fig:pl}}
\end{figure}

\end{widetext}

Absolute mean magnitudes of MW T2Ceps derived in the section \ref{subsec:mwdata} are plotted against the logarithm of periods in Fig. \ref{fig:pl}. We want to find the slope $a$ and intercept $b$ of the PLR:
\begin{equation}
M_{\lambda} = a_{\lambda} (\log P - \log P_0) + b_{\lambda}
\end{equation}
where $P_0$ is the median period of the sample and for MW T2Ceps $\log P_0$=0.3. In order to avoid biases related to inversion of parallaxes, it is suggested to use the Astrometric Based Luminosity (ABL) \citep{1997MNRAS.286L...1F, 1999ASPC..167...13A}. ABL is defined as: 
\begin{equation}
ABL = \pi 10^{0.2 m_{\lambda}-2}=10^{M_{\lambda}/5}
\end{equation}
where $\pi$ is the parallax in miliarcssecods, $m_{\lambda}$ and $M_{\lambda}$ are apparent (corrected for extinction) and absolute magnitudes in the given band $\lambda$, respectively.  We calculate ABL for stars in our sample and parameters $a$ and $b$ with their uncertainties are determined using Monte Carlo simulations and the curvefit routine from the \texttt{Python} \texttt{SciPy} module. We run 10 000 simulations in each passband and in each simulation we generate a random sample of magnitudes and parallaxes from normal distributions defined by mean magnitudes and parallaxes and uncertainties of these values. Slopes and intercepts with corresponding errors are listed in Table \ref{tab:t2cep_plr_mw}.

\begin{deluxetable}{ccccccc}[t!]
\tablenum{4}
\tablecaption{Period-luminosity relations for Milky Way T2Ceps\label{tab:t2cep_plr_mw}}
\tablewidth{0pt}
\tablehead{
\colhead{$Filter$} & \colhead{$a$} & \colhead{$\sigma_a$} & \colhead{$b$} & \colhead{$\sigma_b$} & \colhead{$\sigma$} & \colhead{comment} 
}
\startdata
$J$ & -1.946 & 0.071 & -1.519 & 0.022 & 0.090 & case 1\\
  & -2.005 & 0.084 & -1.525 & 0.023 & 0.062 & case 2\\
  & -1.955 & 0.208 & -1.520 & 0.030 & 0.063 & case 3\\
$H$ & -2.098 & 0.074 & -1.757 & 0.022 & 0.094 & case 1\\
  & -2.173 & 0.088 & -1.766 & 0.023 & 0.041 & case 2\\
  & -2.162 & 0.210 & -1.764 & 0.030 & 0.045 & case 3\\
$K_s$ & -2.140 & 0.075 & -1.827 & 0.023 & 0.103 & case 1\\
  & -2.225 & 0.089 & -1.836 & 0.024 & 0.049 & case 2\\
  & -2.251 & 0.212 & -1.840 & 0.031 & 0.054 & case 3  \\
$W_{JKs}$ & -2.293 & 0.079 & -2.046 & 0.022 & 0.120 & case 1\\
   & -2.399 & 0.095 & -2.057 & 0.022 & 0.059 & case 2\\
  & -2.501 & 0.195 & -2.068 & 0.026 & 0.059 & case 3\\
\enddata
\tablecomments{Case 1- fitted with W Virginis star, case 2- fitted without W Virginis star, case 3- fitted using BL Her stars solely}
\end{deluxetable}

\subsection{LMC}

\begin{widetext}

\begin{figure}[b]
\epsscale{1.}
\plotone{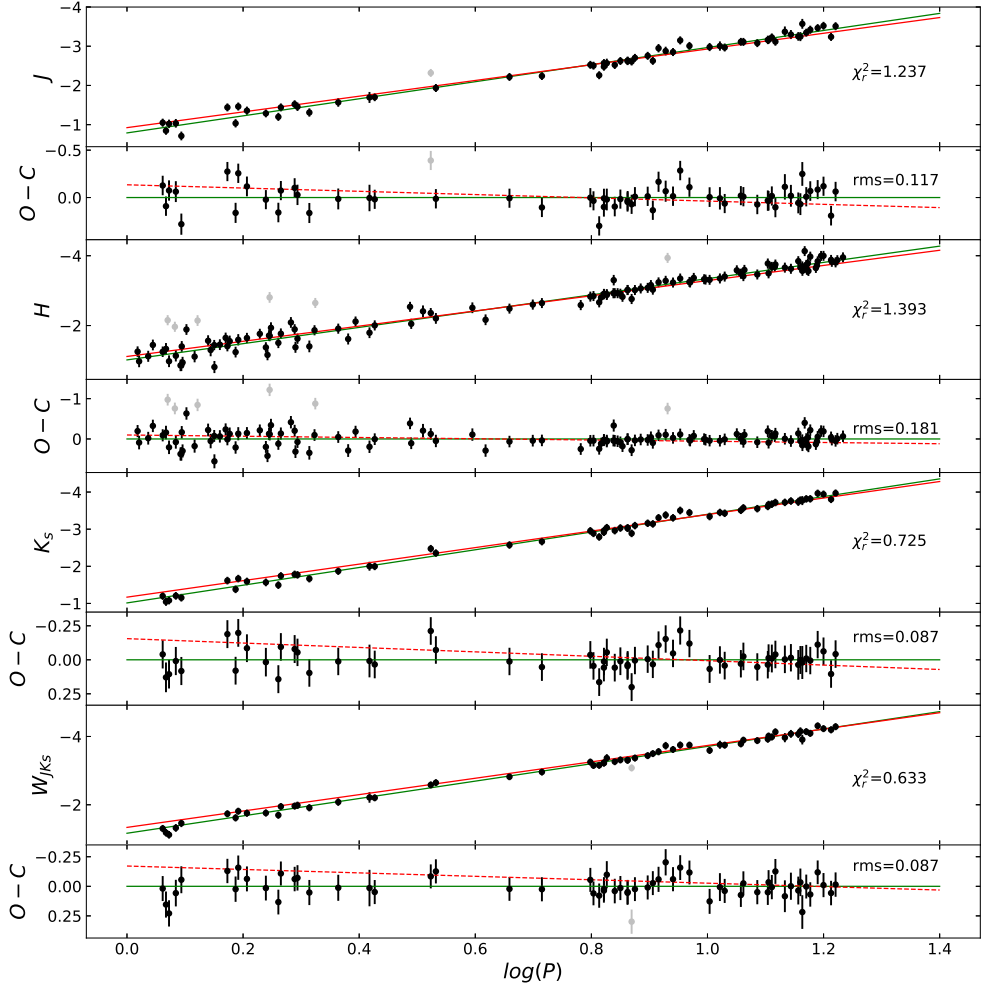}
\caption{Period-luminosity relations of T2Ceps in the Large Magellanic Cloud in $J$, $H$, $K_s$ and $W_{JKs}$ Wesenheit index. Black and grey points mark stars used and excluded from the fitting procedure after 3$\sigma$ clipping applied twice. Green lines are a free fit and the red dashed lines are PLRs obtained for MW T2Ceps (Table \ref{tab:t2cep_plr_mw}). \label{fig:pl_lmc}}
\end{figure}

\end{widetext}

Fig. \ref{fig:pl_lmc} presents the period-luminosity diagrams for LMC T2Ceps. Using the least squares method and 3$\sigma$ clipping (twice) we fit lines in the form of equation 2 (with $\log P_0$=0.7 for the whole sample and $\log P_0$=0.3 when using BL Her stars solely) to obtain slopes and intercepts of these relations. Resulting PLRs for LMC T2Ceps are presented in Table \ref{tab:t2cep_plr_mc} (zero points of PLRs for BL Her+W Vir are recalculated for $\log P_0$=0.3) and plotted in Fig. \ref{fig:pl_lmc} with green lines. Dispersion of the PLR and also the number of 3$\sigma$ outliers is much higher in $H$-band than in other bands, which is expected because for $H$-band we used mean magnitudes inferred from a single measurement. In Fig. \ref{fig:pl_lmc} we also plot PLRs of MW T2Ceps obtained in the previous section and we notice that PLRs are steeper in the LMC in each passband, with a difference greater than $2\sigma$. The observed spread of LMC PLRs is significantly greater than for MW T2Ceps, which clearly shows that distances of MW T2Ceps are determined with a very high precision.

Forcing the MW slopes for PLRs of T2Ceps in the LMC we can measure the distance to the LMC. We use mean apparent magnitudes of LMC T2Ceps corrected for extinction and for the LMC geometry. We fit lines in the form of equation 2 (using $\log P_0$ as above) with slope fixed on the corresponding value from MW relations using least squares method with 3$\sigma$ clipping applied twice. We repeat this fitting using BL Her stars solely and forcing the slope to the corresponding value from MW BL Her stars. Coefficients of fitted lines are listed in Table \ref{tab:lmc_dist}. In column 7 of this table we give the obtained distance modulus $\mu$ of the LMC in each band. The result obtained from $LMCSS$ photometry is more than 0.2mag smaller than in other bands, which was already mentioned in section \ref{subsec:mcdata}. The obtained values in different bands generally agree withing 1$\sigma$, but we note that they are smaller in $J$ band than in $K_s$ band, and are highest in $W_{JKs}$ Wesenheit index, which may be the result of some remaining color excess (underestimated or overestimated extinction towards MW and/or LMC T2Ceps). Results obtained with W Vir and BL Her stars are also in a good agreement (1-2$\sigma$) with the most precise value from eclipsing binaries (18.477), while using BL Her stars solely gives slightly higher distances.

\begin{deluxetable*}{cccccccc}
\tablenum{5}
\tablecaption{Period-Luminosity relations for T2Ceps in the Large Magellanic Cloud. \label{tab:t2cep_plr_mc}}
\tablewidth{0pt}
\tablehead{
\colhead{$Filter$} & \colhead{$a$} & \colhead{$\sigma_a$} & \colhead{$b$@$\log P_0$=0.3} & \colhead{$\sigma_b$} & \colhead{$\sigma$} &  \colhead{$N$} & \colhead{comment}  
}
\decimalcolnumbers
\startdata
$J$ & -2.177 & 0.040 & -1.442 & 0.015 & 0.117 & 61 & BLH+WV\\
  & -2.356 & 0.259 & -1.479 & 0.039 & 0.164 & 20 & BLH\\
$H$ & -2.328 & 0.040 & -1.714 & 0.016 & 0.181 & 122 & IRSF data, BLH+WV\\
  & -2.195 & 0.300 & -1.783 & 0.049 & 0.326 & 54 & IRSF data, BLH\\
  & -1.906 & 0.069 & -2.120 & 0.027 & 0.170 & 39 & LMCSS data, BLH+WV\\
$K_s$ & -2.387 & 0.030 & -1.729 & 0.011 & 0.087 & 62  & BLH+WV\\
  & -2.616 & 0.165 & -1.751 & 0.025 & 0.105 & 20 & BLH\\
$W_{JKs}$ & -2.544 & 0.029 & -1.929 & 0.011 & 0.087 & 61 & BLH+WV\\
  & -2.796 & 0.153 & -1.946 & 0.023 & 0.097 & 20 & BLH\\
\enddata

\tablecomments{BLH, and WV are BL Her and W Vir type stars, respectively}
\end{deluxetable*}

\begin{deluxetable*}{ccccccccc}
\tablenum{6}
\tablecaption{Results of fitting PLRs for LMC T2Ceps with slope fixed on the corresponding value determined from MW sample. \label{tab:lmc_dist}}
\tablewidth{0pt}
\tablehead{
\colhead{$Filter$} & \colhead{$a$} & \colhead{$b$@$\log P_0$=0.3} & \colhead{$\sigma_b$} & \colhead{$\sigma$} &  \colhead{$N$} & $\mu$ & $\sigma_\mu$ & \colhead{comment}  
}
\decimalcolnumbers
\startdata
$J$ & -2.005 & 16.956 & 0.017 & 0.132 & 61 & 18.481 & 0.029 & BLH+WV\\
  & -1.955 & 17.016 & 0.038 & 0.166 & 20 & 18.536 & 0.048 & BLH\\
$H$ & -2.173 & 16.703 & 0.017 & 0.190 & 122 & 18.469 & 0.027 & IRSF data, BLH+WV\\
  & -2.162 & 16.697 & 0.044 & 0.320 & 54 & 18.461 & 0.048 & IRSF data, BLH\\
  & -2.173 & 16.473 & 0.032 & 0.194 & 39 & 18.237 & 0.039 & LMCSS data, BLH+WV\\
$K_s$ & -2.225 & 16.675 & 0.013 & 0.105 & 62 & 18.511 & 0.027 & BLH+WV\\
  & -2.251 & 16.743 & 0.026 & 0.112 & 20 & 18.583 & 0.040 & BLH\\
$W_{JKs}$ & -2.399 & 16.483 & 0.013 & 0.101 & 61 & 18.540 & 0.026 & BLH+WV\\
  & -2.501 & 16.545 & 0.023 & 0.101 & 20 & 18.613 & 0.035 & BLH\\
\enddata

\tablecomments{BLH, and WV are BL Her and W Vir type stars, respectively}
\end{deluxetable*}

\clearpage
\subsection{The Metallicity Effect} \label{subsec:metal_eff}

The last column of Table \ref{tab:t2cep_data} contains spectroscopic metallicity determinations [Fe/H] from \citet{2007ApJ...666..378M} for a fraction of T2Ceps from our MW sample. Uncertainties of these values are not given so we assume $\sigma_{Fe/H}$=0.1dex, which is a rather typical value for measurements of this parameter. Using these measurements and the determined PLRs we investigate a possible metallicity effect on the absolute brightness of T2Ceps. 

\begin{figure}[h]
\epsscale{1.2}
\plotone{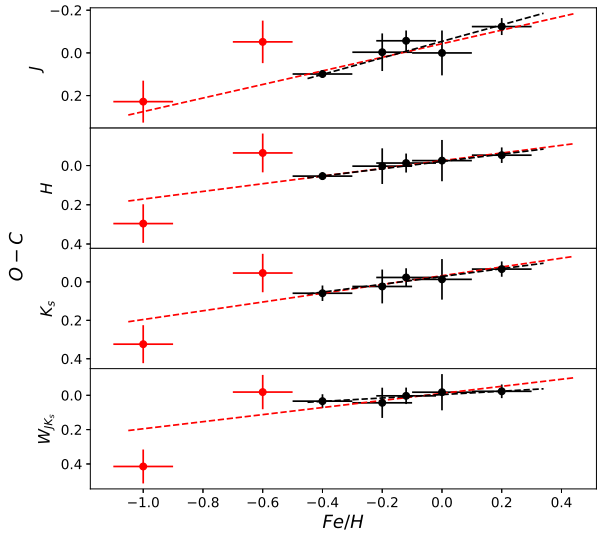}
\caption{Deviation from the period-luminosity relation versus metallicity for Milky Way T2Ceps. Black points marks BL Her stars, red points are W Vir stars. The red line is the fit for W Vir and BL Her stars and the black line for BL Her stars solely. \label{fig:metal_eff}}
\end{figure}

In Fig. \ref{fig:metal_eff} we plot deviations of stars with known [Fe/H] from the PLRs determined in the previous section (in case 1) as a function of [Fe/H]. We continue to reject pW Vir stars and those with low quality parallaxes (TX Del) and also AU Peg. Only 10 stars have metallicities available and 3 of them are rejected so our determination is hampered by poor statistics, however, a pretty clear trend is visible in each band. Using the least squares method (uncertainties are taken into account) we fit lines to these relations in 10000 Monte Carlo simulations, and the slopes of these fits yield an estimation of the $\gamma$ parameter i.e. the influence of the metallicity on the T2Ceps absolute brightness. We fit lines in two cases: a common line for W Vir and BL Her stars (red line) and for BL Her stars solely (black line). Results of this fitting ($\gamma$ values) are presented in Table \ref{tab:t2cep_metaleff}, and they show a significant metallicity effect of the order of -0.2mag/dex in each band and considered case, meaning that more metal-rich T2Ceps are intrinsically brighter than their more metal-poor counterparts. This result is in agreement with the value of -0.1mag/dex obtained by \citet{2006MNRAS.370.1979M} for T2Ceps in Globular Clusters. Recent empirical calibrations of the $\gamma$ parameter for Classical Cepheids gives a very similar value of the effect \citep{2018A&A...620A..99G,2021ApJ...913...38B,2021MNRAS.508.4047R}, while for RR Lyrae stars the sign of the relation is opposite \citep{1994AJ....108..222N,2019MNRAS.490.4254N}.

It is not possible at this point to apply any correction for the metallicity effect to the LMC distance as there are no metallicity determinations for the LMC T2Ceps in the literature and the mean metallicity of our MW sample is not known, thus this effect still contributes to the error of our measurements.


\begin{deluxetable}{cccc}[h]
\tablenum{7}
\tablecaption{Estimation of the metallicity effect on the absolute magnitudes of T2Ceps \label{tab:t2cep_metaleff}}
\tablewidth{0pt}
\tablehead{
\colhead{$Filter$} & \colhead{$\gamma$} & \colhead{$\sigma_{\gamma}$} & \colhead{comment}
}
\startdata
$J$ & -0.318 & 0.074 & BLH+WV\\
  & -0.387 & 0.123 & BLH\\
$H$ & -0.196 & 0.070 & BLH+WV\\
  & -0.186 & 0.090 & BLH\\
$K_s$ & -0.228 & 0.076 & BLH+WV\\
  & -0.203 & 0.111 & BLH\\
$W_{JKs}$ & -0.181 & 0.076 & BLH+WV\\
  & -0.107 & 0.101 & BLH\\
\enddata
\tablecomments{BLH and WV are BL Her and W Virtype stars, respectively}
\end{deluxetable}

\section{Discussion} \label{sec:disc}

\subsection{PLRs} \label{subsec:disc_mw}

Regarding the PLRs of the MW T2Ceps, we notice a very small dispersion for short period BL Her stars which confirm that they are precise distance indicators. Another observation is that on average BL Her stars in the MW are brighter than in the LMC while for W Vir stars the situation is the opposite (see Fig. \ref{fig:pl_lmc}) and the difference between slopes of PLRs is at the level of 2-3$\sigma$. In light of found metallicity effect (assuming that this effect is identical for BL Her and W Vir stars), this could suggest that BL Her and W Vir stars have different mean metallcities. This would in turn support the conclusion of \citet{2018AcA....68..213I} deduced from the spatial distribution of these two groups in the LMC that BL Her and W Vir stars are not the same population of stars. In Table \ref{tab:t2cep_plr_mc_ref} we listed values of slopes of PLRs for T2Ceps in the LMC, GGCs and the GB found in the literature. A general trend of increasing slope towards longer wavelengths is observed in each sample which is in agreement with theoretical studies \citep[e.g.][]{2021MNRAS.501..875D}. PLRs of T2Ceps in the LMC derived in this work are in a very good agreement with other studies of PLRs in this galaxy. MW PLRs are flatter than in other T2Ceps samples and in most cases they differ from other listed slopes by more than $2\sigma$. A very good agreement is observed between our MW PLRs and LMC PLRs from \citet{2017AJ....153..154B}. \citet{2017AJ....153..154B} used mixed $LMCSS$ and $VMC$ photometry but, as we show in our analysis, $LMCSS$ photometry might be problematic.

Having two independent zero-points of PLRs one can average them, but as the slopes of MW and LMC PLRs are quite different and the influence of metallicity on absolute magnitudes might be strong and metallicity determinations of T2Ceps are very poor, we doubt that averaging would increase the precision of distances measured with such averaged PLRs. For the time being we suggest using a single anchor for the T2Ceps distance scale. Future determinations of metallicities of T2Ceps in both samples could allow for studies of a common period-luminosity-metallicity relationship.

An important issue to be discussed is the influence of the $Gaia$ parallaxes ZPO on our MW PLRs. We calculated ZPO using \citet{2021A&A...649A...4L}, later called L21, from $G$ magnitude, color index and ecliptic latitude of a given star, although recently \citet{2021arXiv210608128G} (G21) published their new estimation of the ZPO. Both L21 and G21 used quasars and detached binaries in their calibrations but different selection criteria for these objects. Following the procedure given in G21 we calculated ZPO for our stars. Table \ref{tab:t2cep_g21_zpo} contains $Gaia$ IDs of our stars, HEALPix level giving the best quality of the ZPO \citep[this parameter defines resolution of the used ZPO map, for details see][]{2021arXiv210608128G} and obtained ZPO with errors, which are usually greater than values obtained using L21. We use these corrections to determine PLRs again (excluding W Virginis, which is outlying also with G21 corrections). We also fit PLRs without introducing any ZPO corrections. The results are shown in Table \ref{tab:t2cep_plr_plxzpt}. Not introducing ZPO does not affect slopes while using G21 corrections gives shallower slopes of the PLRs, and the difference is slightly more than 1$\sigma$. Zero-points of PLRs differ by $\sim$0.04mag when using G21, while without ZPO they are about 0.08mag smaller than with ZPO from L21. The influence of using different ZPO is better visible when applying determined PLRs for the LMC distance measurement. Column 7 of Table \ref{tab:t2cep_plr_plxzpt} contains measurements of the distance modulus of the LMC determined using each PLR. The distance modulus obtained without introducing ZPO is by $\sim$0.08mag higher than with L21 correction and much higher than the reference 18.477mag value from eclipsing binaries. G21 corrections gives $\mu$ values lower by $\sim$0.08mag than L21 and also lower than the reference value.

In our calibrations of PLRs for T2Ceps in the MW we used the ABL approach. It might be interesting to see the difference between ABL and a linear regression fit. We repeat fitting in 10000 Monte Carlo simulations, but this time we draw magnitudes in a given band and parallaxes from their respective distributions, and using parallaxes we calculate distance moduli and finally absolute magnitudes of T2Ceps. Then we fit a straight line (equation 2, $\log P_0$=0.3) using the least squares method. Results of this fitting (for case 2 only, i.e. without W Vir star) are also presented in Table \ref{tab:t2cep_g21_zpo} (L.R.). The difference between PLRs obtained with ABL and linear regression is not significant (both slopes and intercepts agree within 1$\sigma$) and the determined distance modulus of the LMC is almost identical.

In Table \ref{tab:t2cep_plr_sys} we summarize systematic errors on zero-points of PLRs determined for the MW and LMC samples. These values are discussed in Section \ref{sec:data}. $\sigma_{tot}$ is the total systematic uncertainty (quadratic sum). The main source of uncertainty in both the MW and the LMC are distances of T2Ceps \citep[$Gaia$ parallaxes and the LMC distance from][]{2019Natur.567..200P}.

\begin{deluxetable*}{cccccccc}[t!]
\tablenum{8}
\tablecaption{Selected literature determinations of the slope of Period-Luminosity Relations for T2Ceps. Sources are: TW is this work, B17a \citep{2017AJ....153..154B}, B17b \citep{2017AA...605A.100B}, R15 \citep{2015MNRAS.446.3034R}, M09 \citep{2009MNRAS.397..933M} and M06 \citep{2006MNRAS.370.1979M}. \label{tab:t2cep_plr_mc_ref}}
\tablewidth{0pt}
\tablehead{
\colhead{$Filter$} & \colhead{host} & \colhead{$a$} & \colhead{$\sigma_a$} & \colhead{$\sigma$} & \colhead{N} & \colhead{source} & \colhead{comment}  
}
\startdata
$J$ & MW & -2.005 & 0.084 & 0.062 & 14 & TW & BLH+WV\\
  & LMC & -2.177 & 0.040 & 0.164 & 61 & TW & BLH+WV\\
  & LMC & -2.061 & 0.038 & 0.157 & 126 & B17a & BLH+WV\\
  & LMC & -2.190 & 0.040 & 0.130 & 120 & R15 & BLH+WV\\
  & GB & -2.240	& 0.031 & 0.316 & 203 & B17b & BLH+WV\\
  & GGCs & -2.230 & 0.050 & 0.160 & 46 & M06 & BLH+WV+RVT\\
$H$ & MW & -2.173 & 0.088 & 0.041 & 14 & TW & BLH+WV\\
  & LMC & -2.328 & 0.040 & 0.181 & 122 & TW & BLH+WV\\
  & LMC & -2.202 & 0.046 & 0.171 & 117 & B17a & BLH+WV\\
  & LMC & -2.316 & 0.043 & 0.200 & 136 & M09 & BLH+WV\\
  & GB & -2.591	& 0.163 & 0.353 & 104 & B17b & BLH\\ 
  & GGCs & -2.340 & 0.050 & 0.150 & 46 & M06 & BLH+WV+RVT\\
$K_s$ & MW & -2.225 & 0.089 & 0.049 & 14 & TW & BLH+WV\\
  & LMC & -2.387 & 0.030 & 0.087 & 62 & TW & BLH+WV\\
  & LMC & -2.232 & 0.037 & 0.180 & 119 & B17a & BLH+WV\\
  & LMC & -2.385 & 0.030 & 0.090 & 120 & R15 & BLH+WV\\
  & GB & -2.189	& 0.056 & 0.234 & 201 & B17b & BLH+WV\\
  & GGCs & -2.410 & 0.050 & 0.140 & 46 & M06 & BLH+WV+RVT\\
$W_{JK_s}$ & MW & -2.399 & 0.095 & 0.059 & 15 & TW & BLH+WV\\
  & LMC & -2.544 & 0.029 & 0.087 & 61 & TW & BLH+WV\\
  & LMC & -2.346 & 0.051 & 0.216 & 119 & B17a & BLH+WV\\
  & LMC & -2.520 & 0.030 & 0.085 & 120 & R15 & BLH+WV\\
\enddata
\tablecomments{BLH, WV and RVT are BL Her, W Vir and RV Tau type stars, respectively}
\end{deluxetable*}

\begin{deluxetable*}{cccccc}
\tablenum{9}
\tablecaption{Parallax zero point offset (ZPO) determined using the procedure from \citet{2021arXiv210608128G}. ZPO$_{L21}$ is the ZPO from L21 listed for comparison.\label{tab:t2cep_g21_zpo}}
\tablewidth{0pt}
\tablehead{
\colhead{Name} & \colhead{$Gaia$ ID} & \colhead{HEALPix level$^1$} & \colhead{ZPO} & \colhead{$\sigma_{ZPO}$}  & \colhead{ZPO$_{L21}$}\\
\colhead{} & \colhead{} & \colhead{} & \colhead{(mas)} & \colhead{(mas)} & \colhead{(mas)} \\
}
\decimalcolnumbers
\startdata
BX Del & 1816085861226864768 & 0 & -0.021 & 0.003 & -0.0188\\
VY Pyx & 5653136461526964224 & 0 & -0.033 & 0.003 & -0.0237\\
BL Her & 4527596850906132352 & 1 & -0.021 & 0.004 & 0.0016\\
KZ Cen & 6144045107427693824 & 0 & -0.011 & 0.003 & -0.0224\\
SW Tau & 3283721030024735360 & 0 & -0.038 & 0.007 & -0.0103\\
V971 Aql & 4188140876549643008 & 0 & -0.024 & 0.003 & -0.0175\\
DU Ara & 5814122315506225792 & 2 & -0.038 & 0.007 & -0.0189\\
V439 Oph & 4472449191647245184 & 1 & -0.029 & 0.004 & -0.0103\\
RT TrA & 5828480459918679936 & 1 & -0.031 & 0.005 & -0.0021\\
V553 Cen & 6217308590845895680 & 0 & -0.017 & 0.003 & -0.0080\\
AU Peg & 1785352625740690432 & 2 & -0.028 & 0.005 & -0.0180\\
FM Del & 1811618408045800832 & 0 & -0.018 & 0.003 & -0.0174\\
TX Del & 1734124248699204096 & 0 & -0.023 & 0.003 & -0.0112\\
$\kappa$ Pav & 6434564460631076864 & 1 & -0.014$^2$ & 0.003 & 0.0046\\
AL Vir & 6303152720661307648 & 0 & -0.016 & 0.003 & -0.0152\\
AP Her & 4510925780739110272 & 1 & -0.021 & 0.003 & -0.0028\\
CO Pup & 5643564972301150208 & 0 & -0.032 & 0.003 & -0.0212\\
W Vir & 3637042116582796544 & 1 & -0.026 & 0.003 & -0.0227\\
ST Pup & 5577329081864722176 & 2 & -0.035 & 0.003 & -0.0091\\
MR Ara & 5954403987593491584 & 0 & -0.027 & 0.003 & -0.0080\\
RS Pav & 6647640365167706240 & 2 & -0.030 & 0.006 & -0.0003\\
\hline
\textbf{mean} & & & \textbf{-0.026} & & \textbf{-0.013}\\
\enddata
\tablecomments{$^1$ HEALPix level defines the resolution of the used ZPO map \citep[for details see][]{2021arXiv210608128G}, $^2$ spatial correction only.}
\end{deluxetable*}

\begin{deluxetable*}{ccccccccc}
\tablenum{10}
\tablecaption{Period-Luminosity relations for Milky Way T2Ceps determined using L21 and G21 parallax zero point offsets (ZPO), and without introducing ZPO (N.A). L.R is the linear regression fit (in 10000 Monte Carlo simulations) using L21 ZPO.\label{tab:t2cep_plr_plxzpt}}
\tablewidth{0pt}
\tablehead{
\colhead{$Filter$} & \colhead{$a$} & \colhead{$\sigma_a$} & \colhead{$b$} & \colhead{$\sigma_b$} & \colhead{$\sigma$} & \colhead{$\mu$} & \colhead{$\sigma_{\mu}$} & \colhead{comment} 
}
\decimalcolnumbers
\startdata
$J$ & -2.005 & 0.084 & -1.525 & 0.023 & 0.062 & 18.481 & 0.029 & L21\\
  & -1.917 & 0.079 &-1.485 & 0.023 & 0.059 & 18.401 & 0.030 & G21\\
  & -2.029 & 0.086 & -1.589 & 0.024 & 0.064 & 18.557 & 0.029 & N.A\\
  & -2.019 & 0.090 & -1.527 & 0.023 & 0.067 & 18.481 & 0.029 & L.R.\\
$H$ & -2.173 & 0.088 & -1.757 & 0.022 & 0.041 & 18.469 & 0.027 & L21\\
  & -2.087 & 0.084 & -1.725 & 0.023 & 0.045 & 18.390 & 0.030 & G21\\
  & -2.198 & 0.091 & -1.829 & 0.024 & 0.057 & 18.536 & 0.030 & N.A\\
  & -2.176 & 0.091 & -1.759 & 0.022 & 0.044 & 18.462 & 0.027 & L.R.\\
$K_s$ & -2.225 & 0.089 & -1.836 & 0.024 & 0.065 & 18.511 & 0.027 & L21\\
  & -2.137 & 0.084 & -1.796 & 0.023 & 0.045 & 18.431 & 0.028 & G21\\
  & -2.250 & 0.090 & -1.901 & 0.024 & 0.052 & 18.587 & 0.027 & N.A\\
  & -2.215 & 0.089 & -1.838 & 0.023 & 0.053 & 18.511 & 0.028 & L.R.\\
$W_{JKs}$ & -2.399 & 0.095 & -2.057 & 0.022 & 0.061 & 18.540 & 0.026 & L21\\
  & -2.315 & 0.090 & -2.019 & 0.022 & 0.061 & 18.464 & 0.027 & G21\\
  & -2.416 & 0.095 & -2.118 & 0.023 & 0.057 & 18.609 & 0.026 & N.A\\
  & -2.351 & 0.093 & -2.059 & 0.024 & 0.057 & 18.520 & 0.025 & L.R.\\
\enddata
\end{deluxetable*}

\begin{deluxetable*}{cccccccccc}
\tablenum{11}
\tablecaption{Summary of the estimated systematic uncertainties of zero-points of PLRs for MW and LMC T2Ceps\label{tab:t2cep_plr_sys}}
\tablewidth{0pt}

\tablehead{\colhead{} & \multicolumn{4}{c}{$MW$} & \colhead{} & \multicolumn{4}{c}{$LMC$} \\
\cline{2-5} 
\cline{7-10}
\colhead{$Filter$} & \colhead{$\sigma _{phot}$} & \colhead{$\sigma _{dist}$} & \colhead{$\sigma _{redd}$} & \colhead{\textbf{$\sigma _{tot}$}} & \colhead{} & \colhead{$\sigma _{phot}$} & \colhead{$\sigma _{dist} $} & \colhead{$\sigma _{redd}$} & \colhead{\textbf{$\sigma _{tot}$}}
}
\startdata
$J$ & 0.002 & 0.024 & 0.02 & \textbf{0.031} & & 0.02 & 0.026 & 0.01 & \textbf{0.034}\\
$H$ & 0.002 & 0.024 & 0.02 & \textbf{0.031} & & 0.05 & 0.026 & 0.01 & \textbf{0.057}\\
$K_s$ & 0.002 & 0.024 & 0.01 & \textbf{0.026} & & 0.02 & 0.026 & 0.01 & \textbf{0.034}\\
$W_{JKs}$ & 0.003 & 0.024 & 0.01 & \textbf{0.026} & & 0.03 & 0.026 & 0.01 & \textbf{0.041}\\
\enddata
\tablecomments{$\sigma _{phot}$ is the photometric zero-point uncertainty; $\sigma _{dist}$ is the distance error; $\sigma _{redd}$ is the reddening error;  $\sigma _{tot}$ is total systematic uncertainty (quadratic sum)}
\end{deluxetable*}

\subsection{The LMC distance} \label{subsec:disc_lmc}

In column 8 of Table \ref{tab:t2cep_plr_mc} we give statistical errors of our LMC distance modulus measurements and Table \ref{tab:t2cep_plr_sys} presents sources of systematic uncertainty on both MW and LMC sample. Systematic uncertainty on the LMC distance modulus in each band is quadratic sum of $\sigma _{tot}$ for the MW sample and $\sigma _{phot}$ and $\sigma _{redd}$ for the LMC sample. It gives us the following values. For $J$ and $K_s$ bands we obtain 0.038mag, 0.060mag for $H$ band and 0.034mag for $W_{JKs}$ Wesenheit index. We emphasize that this values are most probably underestimated as we do not take into account the effect of metallicity in the distance determination. As we describe in section \ref{subsec:disc_mw}, using ZPO for $Gaia$ parallaxes from G21 instead of L21 give us quite discrepant results, thus the ZPO uncertainty stated by \citet{2021A&A...649A...4L} can be underestimated.

Previous determinations of the LMC distance based on T2Ceps rely on the Hipparcos parallax measurements of 2 stars ($\kappa$ Pav and VY Pyx) and pulsational parallaxes of $\kappa$ Pav, V553 Cen and SW Tau \citep{2008MNRAS.386.2115F}. Excluding pW Vir star $\kappa$ Pav, the absolute distance scale of T2Ceps was tied to 3 stars only. Using pulsational parallaxes of V553 Cen and SW Tau \citet{2009MNRAS.397..933M} found the LMC distance modulus to be 18.46$\pm$0.10 mag. \citet{2015MNRAS.446.3034R} in their Fig. 11 show measurements of the distance modulus of the LMC based on different MW calibrators. They used four stars mentioned above (including $\kappa$ Pav) in addition to RR Lyrae stars with $HST$ parallaxes from \citet{2011AJ....142..187B} and from different combinations of these anchors they obtained a value of about 18.57mag.

The most accurate distance to the LMC was measured using late-type eclipsing binaries by \citet{2019Natur.567..200P} and they obtained the distance modulus of 18.477$\pm$0.004(stat.)$\pm$0.026(syst.)mag. The results obtained in this work using BL Her and W Vir stars agrees within 1$\sigma$ with this value. Using BL Her stars solely gives significantly higher results. This may be the result of the metallicity effect and/or slightly different geometrical distribution of these stars in the LMC compared to eclipsing binaries from \citet{2019Natur.567..200P} and W Vir stars, as well as lower quality of the photometry of these stars in the LMC due to their low brightness and crowding.

\subsection{Metallicity Effect} \label{subsec:disc_metal_eff}

We investigated the metallicity effect using ZPO from L21, thus we repeat determination of the $\gamma$ parameter using G21 ZPO for $Gaia$ parallaxes and not introducing any ZPO and we obtain almost identical results. Obtained values of the $\gamma$ parameter are in most cases higher than -0.1mag/dex found by \citet{2006MNRAS.370.1979M} but we can confirm the sign of the effect. It is important to mention that metal abundances of our sample stars are much higher than those in \citet{2006MNRAS.370.1979M} from GGCs, and that the most metal poor stars in our sample are two W Vir type stars. Very similar values of $\gamma$ have been obtained recently empirically for CCeps \citep{2018A&A...620A..99G,2021ApJ...913...38B,2021MNRAS.508.4047R}. On the other hand, theoretical studies by \citet{2021MNRAS.501..875D} and \citet{2007A&A...471..893D} yield the metallicity effect for T2Ceps of opposite sign and actually predict a null effect in the near-infrared regime. The metallicity effect having opposite sign is also observed for RR Lyrae stars \citep[e.g. ][]{1994AJ....108..222N,2019MNRAS.490.4254N}. Determinig metal abundances for the whole sample of the field T2Ceps in a homogeneous way using high resolution spectra is crucial for a more precise investigation of the metallicity effect.

\section{Summary and Conclusions} \label{sec:concl}
Using new photometric data and $Gaia$ EDR3 parallaxes of the field T2Ceps we determined the most precise PLRs for these radially pulsating stars in the near infrared $J$, $H$, $K_s$ bands and $W_{JK}$ Wesenheit index. We redetermine T2Cep PLRs in the LMC using archival photometry and the most precise distance to this galaxy from eclipsing binaries. Slopes of the obtained MW and LMC PLRs, as well as PLRs for Galactic Globular Clusters and Galactic Bulge from the literature, agree at $2\sigma$ level. We used the obtained MW PLRs to measure the distance modulus of the LMC, and our result in the Wesenheit index $W_{JKs}$ is 18.540$\pm$0.026(stat.)$\pm$0.034(syst.)mag. Using literature values of metal abundanes available for a fraction of field T2Ceps we investigated the metallicity effect and find it to be of the order of -0.2mag/dex, in agreement with the value obtained from studies of T2Ceps in GGCs.

\begin{figure}
\epsscale{1.2}
\plotone{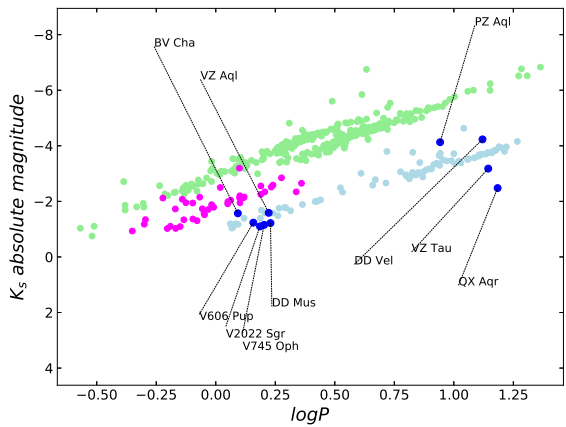}
\caption{The $K_s$ band period-luminosity diagram for nearby stars catalogued as T2Ceps and not used in this study (blue points) based on the 2MASS photometry. Absolute magnitudes are calculated from $Gaia$ EDR3 parallaxes. The LMC T2Ceps (cyan), CCeps (green) and Anomalous Cepheids (magenta) from the VMC catalogue are plotted for comparison.\label{fig:addt2cep}}
\end{figure}

In Fig. \ref{fig:addt2cep} we present a period-luminosity diagram for LMC T2Ceps, CCeps and Anomalous Cepheids and nearby pulsating stars not used in this study but classified as T2Ceps in the literature which are located well on the PLR determined by the LMC sample (and visible from OCA). We plan to collect near-infrared photometry for these stars in the future to increase precision of fiducial PLRs. Multiband analysis would also benefit from observations in additional passbands thus we plan to observe these stars in the optical regime. As we mentioned before, precise determination of the metallicity effect for T2Ceps requires homogeneous metal abundances for the whole analysed sample thus we have started to collect high-resolution spectra with this end in view.

The main source of systematic error on our PLRs and the LMC distance determination is the $Gaia$ parallax zero point. We believe that future investigations into this parameter will significantly improve precision and accuracy of PLRs calibrations as well as distance measurements using T2Ceps.


\begin{center}
ACKNOWLEDGMENTS
\end{center}

We thank the referee for very valuable comments and suggestions which helped us to improve the manuscript.

P.W. gratefully acknowledges financial support from the Polish National Science Center grant PRELUDIUM 2018/31/N/ST9/02742. The  research  leading  to  these  results  has  received  funding from the European Research Council (ERC) under the European  Union's  Horizon  2020  research  and  innovation  programme  under  grant  agreement  No 695099 (project  CepBin). Support from  DIR/WK/2018/12  grant  of the Polish Ministry of Science and Higher Education and the Polish National Science Center grants MAESTRO 2017/26/A/ST9/00446 and BEETHOVEN 2018/31/G/ST9/03050 is also acknowledged. W.G. and G.P. gratefully acknowledge financial support for this work from the BASAL Centro de Astrofisica y Tecnologias Afines (CATA) AFB-170002. B.P. gratefully acknowledges support from the Polish National Science Center grant SONATA BIS 2020/38/E/ST9/00486. R.S. gratefully acknowledges support from the Polish National Science Center grant SONATA BIS 2018/30/E/ST9/00598. A.G acknowledges support from the ALMA-ANID fund No.ASTRO20-0059. F.P gratefully acknowledges the generous and invaluable support of the Klaus Tschira Foundation.

Based on data collected under the ESO/CAMK PAN – USB agreement at the ESO Paranal Observatory.

This work has made use of data from the European Space Agency (ESA) mission
{\it Gaia} (\url{https://www.cosmos.esa.int/gaia}), processed by the {\it Gaia}
Data Processing and Analysis Consortium (DPAC,
\url{https://www.cosmos.esa.int/web/gaia/dpac/consortium}). Funding for the DPAC
has been provided by national institutions, in particular the institutions
participating in the {\it Gaia} Multilateral Agreement.

This publication makes use of data products from the Two Micron All Sky Survey, which is a joint project of the University of Massachusetts and the Infrared Processing and Analysis Center/California Institute of Technology, funded by the National Aeronautics and Space Administration and the National Science Foundation. This research has made use of the SIMBAD database, operated at CDS, Strasbourg, France \citep{2000A&AS..143....9W}. We acknowledge with thanks the variable star observations from the AAVSO International Database contributed by observers worldwide and used in this research.

We also thank our colleagues: Miguel Murphy, Marcin G\l{}adkowski, Susanne Blex, Simon Borgniet, Zohreh Ghaffari, Behnam Ghazinouri, Vincent Hocde, Timofej Lisow, Michael Ramolla, Fabian Symietz and Boris Trahin for their great help in observing in Observatorio Cerro Armazones. 

\software{gaiadr3\_zeropoint \citep{2021A&A...649A...4L}, Astropy7 \citep{2013A&A...558A..33A,2018AJ....156..123A}, fnpeaks (\url{http://helas.astro.uni.wroc.pl/deliverables.php?active=fnpeaks}), IRAF \citep{1986SPIE..627..733T,1993ASPC...52..173T}, Sextractor \citep{1996A&AS..117..393B}, SCAMP \citep{2006ASPC..351..112B}, SWARP \citep{2010ascl.soft10068B}, DAOPHOT \citep{1987PASP...99..191S}, NumPy \citep{2011CSE....13b..22V,2020Natur.585..357H}, SciPy \citep{2020SciPy-NMeth}, Matplotlib \citep{2007CSE.....9...90H}}

\newpage

\bibliography{t2cep}{}
\bibliographystyle{aasjournal}



\end{document}